\documentclass[english,aps,prb,showpacs,twocolumn]{revtex4}
\usepackage{amsmath}
\usepackage{graphicx}

\newcommand{\bfE}{ {\bf E} }

\newcommand{\bfr}{ {\bf r} }
\newcommand{\bfp}{ {\bf p} }

\newcommand{\ba}{\begin{array}}
\newcommand{\ea}{\end{array}}

\newcommand{\beq}{\begin{equation}}
\newcommand{\eeq}{\end{equation}}

\newcommand{\bff}{{\mbox{\boldmath $f$}}}
\newcommand{\bfn}{{\mbox{\boldmath $n$}}}

\newcommand{\bfm}{{\mbox{\boldmath $m$}}}
\newcommand{\bfj}{{\mbox{\boldmath $j$}}}
\newcommand{\bfJ}{{\mbox{\boldmath $J$}}}
\newcommand{\bfsigma}{{\mbox{ \boldmath $\sigma$ }}}
\newcommand{\beqa}{\begin{eqnarray}}
\newcommand{\eeqa}{\end{eqnarray}}
\makeatother

\begin{document}

\title{Spin transfer torque in continuous textures:\\
semiclassical Boltzmann approach}

\author{Fr{\'{e}}d{\'{e}}ric Pi{\'{e}}chon and Andr{\'{e}} Thiaville}
\affiliation{Laboratoire de Physique des Solides, B{\^{a}}t. 510, CNRS,
Univ. Paris-sud,
91405 Orsay, France}

\begin{abstract}
We consider a microscopic model of itinerant electrons coupled via
ferromagnetic exchange to a local magnetization
whose direction vector $\bfn(\bfr,t)$ varies in space and time.
We assume that
to first order in the spatial gradients and time derivative of
$\bfn(\bfr,t)$ the magnetization distribution function $\bff(\bfp,\bfr,t)$ of
itinerant electrons with momentum $\bfp$ at position $\bfr$ and time $t$ has the
Ansatz form
$\bff(\bfp,\bfr,t)=f_{\parallel}(\bfp)\bfn(\bfr,t)+
f_{1 \bfr}(\bfp)\bfn \times \nabla_{\bfr} \bfn+
f_{2 \bfr}(\bfp)\nabla_{\bfr} \bfn+
f_{1 t}(\bfp)\bfn \times \partial_t \bfn+
f_{2 t}(\bfp)\partial_t \bfn$.
Using the Landau-Sillin equations of motion approach
we derive explicit forms for the components
$f_{\parallel}(\bfp), f_{1 \bfr}(\bfp), f_{2 \bfr}(\bfp),
f_{1 t}(\bfp)$ and $f_{2 t}(\bfp)$ in "equilibrium" and in out-of-
equilibrium situations for (i) no scattering by impurities,
(ii) spin-conserving scattering and
(iii) spin non-conserving scattering.
The back action on the localized electron magnetization from the out-of-equilibrium
part of the two components $f_{1 \bfr}, f_{2 \bfr}$ constitutes the two
spin transfer torque terms.

\end{abstract}

\maketitle

\section{Introduction}

Recent experiments in spin-valve nanopillars \cite{Katine00, kiselev}, point
contacts \cite{Chen04} and ferromagnetic nanowires
\cite{Vernier04,Yamaguchi04, Klaui05, Hayashi06, Thomas06} have demonstrated
the possibility to "manipulate" the magnetization
by applying an electrical current instead of an external magnetic field.
It is believed that this phenomenon might give rise to many technological applications,
(MRAM, fast magnetic switching, high quality microwave sources),
provided that the current density necessary to manipulate the magnetization can be
drastically reduced.

This possibility of electrical current induced magnetization manipulation was
predicted already ten years ago by J. C. Slonczewski and L. Berger \cite{slonczewski,berger}.

Nowadays, the consensus is that, in presence of an electrical
current, the standard Landau-Lifshitz-Gilbert (LLG) equation that
describes the dynamics of the magnetization is modified.
For ferromagnetic wires in which the magnetization texture
${\bf M}(\bfr,t)=M_s\bfn(\bfr,t)$ has a time and space dependent direction
$\bfn(\bfr,t)$, the modified LLG equation reads \cite{zhangli}:
\beq
\partial_t \bfn=\gamma  {\bf B}_{\textrm{eff}}(\bfn) \times \bfn +
\alpha \bfn \times \partial_t\bfn -
{\bf u} \cdot \nabla_{\bfr} \bfn +
\beta \bfn \times {\bf u} \cdot \nabla_{\bfr} \bfn.
\eeq
where ${\bf B}_{\textrm{eff}}(\bfn)=\frac{-1}{M_s}
\dfrac{\delta{\cal E}(\bfn)}{\delta \bfn}$ is the effective magnetic field derived
from the magnetic free energy
and $u$, $\beta$ are two phenomenological parameters where ${\bf u}$ is proportional
to the electrical current density and polarization, and $\alpha$ is the usual
phenomenological Gilbert damping parameter.
The contribution ${\bf u} \cdot \nabla_{\bfr} \bfn$ is called the adiabatic
term and can be
derived from an additional term in the magnetic free energy that takes into account
the coupling to the electrical current \cite{bazaliy}.
The $\beta$ term is in contrast non adiabatic and appears to play a role
similar to the Gilbert damping term (and one conception of the spin-transfer
torque introduction in the LLG equation even predicts $\beta = \alpha$
exactly \cite{barnes}).
Recent micromagnetic numerical experiments
\cite{thiaville1,thiaville2,lizhang,he} with this
modified LLG equation have greatly clarified the qualitative
roles played by the two phenomenological parameters $u$ and $\beta$.
In particular, it has been proved that in the absence of the $\beta$ term there
is no
current induced steady domain wall motion below a finite (very high) critical current
density $u_c$.
For a non zero $\beta$ term, in the simple case of a perfect sample it can be shown that
the speed of the domain wall is $(\beta/\alpha) u$.
More quantitatively, results compatible with the various
experiments are obtained for a ratio $\beta/\alpha \ge 1$ equivalent to
$\beta \simeq 10^{-2}$.

The derivation of the LLG equation above usually rests on a two
steps argument.
It is assumed that an itinerant ferromagnet can be
modeled as a two "components" system:
(i) non moving and ferromagnetically ordered electrons (called hereafter $d$) described
by a classical magnetization vector $\bfn(\bfr,t)$ that varies slowly with time
$t$ and position $\bfr$;
(ii) current-carrying itinerant electrons (called  $s$) coupled to the $d$ electrons via
a ferromagnetic exchange energy $\Delta_{sd} > 0$ such that the effective one-electron
quantum Hamiltonian has the form
$\hat H(\bfr)=-\frac{\hbar^2}{2m}\nabla_{\bfr}^2 \hat I+
\frac{\Delta_{sd}}{2}\bfn(\bfr,t)\cdot \hat \bfsigma$,
where $\hat I$ is the $2\times 2$ identity matrix and $\hat \bfsigma$ is the
vector of Pauli matrices with eigenvalues $\pm 1$.

The first step consists in finding the quantum average itinerant electron magnetization
$\bfm(\bfr,t)=-\mu_B \text{Tr}\lbrace \hat \rho(\bfn) \hat \bfsigma\rbrace$ as a function
of the "quasistatic" $\bfn(\bfr,t)$ ($\hat \rho(\bfn)$ is the itinerant electron density
matrix that depends on $\bfn(\bfr,t)$).
In a second step one substitutes
 the resulting $\bfm(\bfr,t)$ back into the LLG equation of
the localized electrons magnetization:
\begin{equation}
\label{eq:LLG2}
\partial_t\bfn=\gamma {\bf B}_{\textrm{eff}}(\bfn) \times \bfn +
\alpha \bfn \times \partial_t \bfn +
\frac{\gamma}{M_s \mu_B}\Delta_{sd} \bfm(\bfr,t) \times \bfn.
\end{equation}
For domain walls, Zhang and Li (ZL) \cite{zhangli} were the first to make a transparent
derivation along this line of reasoning.
They found that in out-of-equilibrium situation the above back action not
only produces the two spin transfer torque terms but also leads to correction of the
Gilbert damping term and gyromagnetic ratio.
Prior to ZL, Zhang, Levy and Fert \cite{zhang1} had obtained
corresponding results for spin-valves.
Both works however rest on a phenomenological equation of motion for the magnetization
$\bfm(\bfr,t)$ of itinerant electrons where key ingredients
are put by hand, especially spin flip scattering time and adiabaticity of the
itinerant electron spin current with respect to local magnetization $\bfn(\bfr,t)$.
Beside these phenomenological descriptions, various microscopic
approaches have also flourished in the last years. For spin-valves systems
several groups used a scattering matrix approach (for a review see \cite{review}).
For ferromagnetic wires the concept of local spin reference
frame was often invoked so as to exhibit a direct coupling between
spin current of itinerant electrons and gradient of the local
magnetization $\bfn(\bfr,t)$ \cite{bazaliy,tatara1,waintal,fernandez,barnes,xiao}
To the best of our understanding however, none of these different works
really succeeds to establish the modified LLG equation as written above.

Very recently two independent works, by Tserkovnyak {\it et al.} \cite{tserkovnyak} and
Kohno {\it et al.} \cite{kohno},
based on different theoretical techniques presented a direct microscopic derivation of
the two additional spin torque terms.
Both works show that the $\beta$ term requires the existence of a spin flip like
scattering mechanism (e.g. spin non-conserving scattering like spin orbit, magnetic
impurities ...).
They show that this mechanism is also responsible for the appearance of an effective
$\alpha$ Gilbert damping term induced by the itinerant electrons.
In fact Tserkovnyak {\it et al.} argue further that one should find $\alpha=\beta$ for 
itinerant
ferromagnetic systems where the magnetism comes from the exchange interaction between
the itinerant electrons (e.g. the effective local $\bfn(\bfr,t)$ is in fact the
itinerant electron magnetization  $\bfm(\bfr,t)$ itself).
The results of Kohno {\it et al.} were obtained from diagrammatic linear response theory
and concerned only "integrated" physical quantities like the local magnetization
$\bfm(\bfr,t)$.
In contrast Tserkovnyak {\it et al.} used the Keldysh quasiclassical Green function 
technique that in principle could allow determining the full magnetization distribution 
function $\bff(\bfp,\bfr,t)$, and thus might give a deeper
understanding of the system.


In this paper we reconsider the entire problem within a Boltzmann approach.
This provides an intuitive and hopefully
pedagogical semiclassical picture of the equations of
motion of the charge and spin distribution functions of the itinerant electrons, in a
space-time dependent magnetization field $\bfn(\bfr,t)$.
Our main assumption is that around any time space position the direction of
$\bfn(\bfr,t)$ can be arbitrary but its gradients $\nabla_{\bfr}\bfn$ and
$\partial_t\bfn$ must be slow enough so that only terms parametrically linear
in these gradients are important
(e.g terms like $\nabla_{\bfr}\bfn$, $\partial_t\bfn$,
$\bfn\times \nabla_{\bfr}\bfn$ and $\bfn\times \partial_t\bfn$).
This parametrization is different from that of Kohno {\it et al} and Tserkovnyak 
{\it et al.} which is in fact a linear theory around the uniform magnetization 
case (e.g. $\bfn(\bfr,t)=\hat z +{\bf u}(\bfr,t)$ with $|{\bf u}(\bfr,t)| \ll 1$).
The difference might appear subtle, but it leads to distinct properties
already in the equilibrium situation as compared to Tserkovnyak {\it et al.}.

The Boltzmann method provides a complementary physical picture
that is intermediate between the purely phenomenological
macroscopic equations of motion approach of ZL on one side and the less intuitive 
but more microscopic quantum linear response or Keldysh methods on the other side.
More precisely, on the one hand, since we deal directly
with the charge and spin distribution functions,
we have a deeper understanding than that provided by ZL method.
We have also access to more microscopic physical quantities, and can in principle 
study general AC and thermoelectric effects not accessible to ZL method \cite{inpreparation}.
On the other hand, the physical picture provided by the Boltzmann method is more 
transparent than the microscopic approach, it also allows understanding more clearly 
at which level spin-flip scattering differs qualitatively from spin-conserving scattering.
Lastly, the Boltzmann method provides the natural framework to see where quantum 
corrections could be important.
Indeed, in a forthcoming paper
we will reexamine our results using the Keldysh Green function
method in the quasiclassical approximation. 
This constitutes the natural theoretical
framework to build from first principles the equations of motion for the distribution
function in the presence of elastic and inelastic collisions.
The Keldysh approach appears necessary because the construction of the collision
integral in the Boltzmann picture is purely phenomenological so that, on a more 
microscopic level,
it is not clear if there are important quantum and gradient corrections to
the Boltzmann collision integrals.

The paper is organized as follows.
In the first section, going back to textbooks \cite{sillin,white} we rederive the
collisionless transport equations for the charge and magnetization distribution to
first order in time-space gradients.
We show that without any collision the magnetization distribution
function $\bff(\bfp,\bfr,t)$ is not collinear to $\bfn(\bfr,t)$ and that, in particular,
there is a contribution collinear to $\bfn \times \nabla_{\bfr} \bfn$ that gives 
rise to a finite equilibrium spin current.
In the next two sections with study the influence of elastic scattering by impurities.
Assuming Boltzmann type collision integrals, we first consider the effect of
spin-conserving collisions, in equilibrium and in out-of-equilibrium situation.
Already at equilibrium we obtain surprising results for the  components of the
distribution that are not collinear to $\bfn(\bfr,t)$.
We show in particular that the equilibrium spin current is rotated in the 
direction $\nabla_{\bfr} \bfn$ by an angle $\theta$ that depends on the ratio between 
the elastic scattering time and the effective Larmor time. 
In out-of-equilibrium situation we find correspondingly a component of the magnetic 
distribution function collinear to $\nabla_{\bfr} \bfn$ that could in principle give 
rise to a $\beta$ term as back action after $\bfp$ momentum integration. 
(Un)fortunately it vanishes after $\bfp$ momentum integration as spin conservation 
imposes. 
We then consider collisions that lead to spin flip.
For the case of a uniform magnet, starting from the known form of the collision integral
of each eigen-spin distribution, we build a collision integral invariant under spin basis
change. 
Extending phenomenologically this collision integral to the case of a
non uniform ferromagnet we describe how the equilibrium and out-of-equilibrium properties
are modified by the spin flip scattering. 
In particular, from the different components of the itinerant electron magnetization 
we evaluate their back action on the $d$ electron magnetization and then 
extract explicit expressions for the induced Gilbert damping $\alpha_{2t}$, 
modified adiabatic torque term $u$ and finally the $\beta$ torque term. 
Our expressions for $\alpha_{2t}$ and $\beta$ coincide exactly with ZL approach and to 
leading order also coincide with results of Kohno {\it et al} for spin-isotropic 
spin flip scattering, when appropriate changes of notations are made. 
Concerning the parameter $u$, that is purely phenomenological in the ZL approach, our result
is similar to that of Kohno {\it et al}.
Note finally that, in order to improve reading, most of the calculations concerning 
spin flip scattering are put in the Appendix whereas the main text essentially provides
the physical picture emerging from the modified distribution functions.
In a last concluding section
we discuss possible extensions of our approach
to itinerant ferromagnets or ferromagnetic Fermi liquids \cite{mineev,leggett}
and spin-valve systems \cite{spinvalve}.

\section{Semiclassical transport theory}

\subsection{Model and Ansatz}

The effective one-electron quantum Hamiltonian of the itinerant free electrons
coupled to the localized electron magnetization is of the form
\beq
\hat H(\bfr)=[-\frac{\hbar^2}{2m}\nabla_{\bfr}^2+V(\bfr) ] \hat I+
\frac{\Delta_{sd}}{2}\bfn(\bfr,t)\cdot \hat \bfsigma,
\eeq
where $V(\bfr)=-e\bfE \bfr$ is the potential induced by an external uniform electric field.
The intrinsic difficulty to find the equilibrium and out-of-equilibrium density matrix
$\rho(\bfr,\bfr',t)$ associated to this Hamiltonian originates from the non commutation
of the "Zeeman" term with the kinetic term, due to the spatial variation of $\bfn(\bfr,t)$.

However, as we consider here domain walls where the characteristic length of the
magnetization gradient is large (10-100~nm) compared to the electron mean free path,
quantum transport is not pertinent and electron diffusion is a more appropriate
framework.
Therefore we simplify the problem by assuming that the spatial degrees of freedom
$\bfr,\bfp$ are classical commuting variables and not quantum operators, and retain
only the non trivial commutation rules of spin degrees of freedom.
The effective semiclassical Hamiltonian of the itinerant electrons is thus
\beq
\hat H(\bfr,\bfp)=[\frac{\bfp^2}{2m}+V(\bfr) ] \hat I+
\frac{\hbar \omega_{sd}}{2}\bfn(\bfr,t)\cdot \hat \bfsigma,
\eeq
where we have denoted by $\omega_{sd}=\frac{\Delta_{sd}}{\hbar}$ 
the effective Larmor frequency.
We further define
$\tau_{sd}=\frac{1}{\omega_{sd}}$,
and $\ell_{sd}=v_F \tau_{sd}$ the Larmor time
and Larmor length respectively ($v_F$ is the Fermi velocity).

The semiclassical quantity that corresponds to the density matrix is now the
spinor distribution function
\beq
\label{eq:fdef}
\hat f(\bfp,\bfr,t)=
\frac{1}{2}[f(\bfp,\bfr,t) \hat I+\bff(\bfp,\bfr,t)\cdot \hat \bfsigma].
\eeq
The physical quantities such as local particle density $n(\bfr,t)$, particle
current density $\bfj(\bfr,t)$, magnetization density $\bfm(\bfr,t)$ and
spin-current tensor density $\bfJ(\bfr,t)$ are obtained
by integration on these distributions:
\beq \ba{ll}
n(\bfr,t)&=\int d\tau\textrm{Tr}_{\sigma}\lbrace \hat f(\bfp,\bfr,t)\rbrace\\
&=\int d\tau f(\bfp,\bfr,t),\\
\bfj(\bfr,t)
&=\int d\tau\frac{\bfp}{m}
f(\bfp,\bfr,t)\\
\bfm(\bfr,t)&=-\mu_B \int d\tau\textrm{Tr}_{\sigma}\lbrace \hat f(\bfp,\bfr,t)  
\hat \bfsigma \rbrace\\
&=-\mu_B \int d\tau
\bff(\bfp,\bfr,t)\\
\bfJ(\bfr,t)&=-\mu_B \int d\tau\frac{\bfp}{m}
\bff(\bfp,\bfr,t)
\ea \eeq
where
$$
\int d\tau \equiv \int \dfrac{\textrm{d}\bfp}{(2\pi\hbar)^3}
\equiv \int d\epsilon_{\bfp} \nu(\epsilon_{\bfp}) \int \dfrac{\textrm{d}\hat \bfp}{4\pi}$$
with $\hat \bfp$ the unit vector of direction $\bfp$, $\epsilon_{\bfp}=\frac{\bfp^2}{2m}$ 
and
$\nu(\epsilon)=\dfrac{\sqrt{2m^3 \epsilon}}{2\pi^2 \hbar^3}$
the 3D free electrons density of states.

Our main assumption is that around any space-time position $\bfr,t$, the direction
$\bfn(\bfr,t)$ can be arbitrary but its gradients $\nabla_{\bfr}\bfn$
(resp. $\partial_t\bfn$)
must be slow enough compared to the Larmor length $\ell_{sd}$ (resp.
Larmor time) so that only terms linear in these
gradients are important.
Linearization in these gradients implies that
time and space dependencies of the matrix distribution function
$\hat f(\bfp,\bfr,t)$ are expanded on the possible directions
$\bfn(\bfr,t)$, $\bfn \times \nabla_{\bfr} \bfn$,
$\bfn \times \partial_t \bfn$, $\nabla_{\bfr} \bfn$, and $\partial_t \bfn$.
The Ansatz form we assume for $\hat f(\bfp,\bfr,t)$ compatible with this
approximation is :
\beq 
\ba{ll}
f(\bfp,\bfr,t)=& f(\bfp),\\
\bff(\bfp,\bfr,t)=&f_{\parallel}(\bfp)\bfn(\bfr,t)\\
&+\ell_{sd}\left[ f_{1 \bfr}(\bfp) \bfn \times \nabla_{\bfr} \bfn+
f_{2 \bfr}(\bfp)  \nabla_{\bfr} \bfn \right]\\
&+\tau_{sd} \left[ f_{1 t}(\bfp) \bfn \times \partial_t \bfn +f_{2 t}(\bfp)
\partial_t \bfn \right]. 
\ea 
\eeq
Note that $(\bfn,\partial_t \bfn, \bfn \times \partial_t \bfn)$ and
$(\bfn,\nabla_{\bfr} \bfn, \bfn\times \nabla_{\bfr} \bfn)$ constitute
two distinct orthogonal bases for any spin vector.
Thus {\it a priori} our Ansatz contains some redundancy since
$\partial_t \bfn$, $\bfn \times \partial_t \bfn$ are linear
combinations of $\nabla_{\bfr} \bfn$, $\bfn\times \nabla_{\bfr} \bfn$ and
reciprocally.
The main reason why
our Ansatz is nevertheless appropriate is that, in this extended
basis $(\bfn,\partial_t \bfn, \bfn \times \partial_t
\bfn,\nabla_{\bfr} \bfn, \bfn\times \nabla_{\bfr} \bfn)$, to leading order,
each component of the spin distribution is stationary (for DC field), 
space independent, and depends only on $\bfp$ \cite{footnote1}.
Had we chosen the
basis $(\bfn,\partial_t \bfn, \bfn \times \partial_t \bfn)$ to expand
the spin distribution, each component would still be
position-dependent to leading order.

Note that with our normalization the components $f_{1
\bfr}$, $f_{2 \bfr}$, $f_{1 t}$ and $f_{2 t}$ do have the same
physical dimension as $f_{\parallel}$. 
As a consequence, the quantity
$m_{1\bfr }(\bfr,t)=-\mu_B \int d\tau f_{1\bfr}(\bfp,\bfr,t)$ has the units of
a magnetization density.

Following Landau-Sillin \cite{sillin}, to first order in the gradients of the
distribution
function, the "Liouville" equation of motion of the distribution is obtained from
\beq
\frac{d\hat f}{dt} \equiv
\partial_t \hat f +
\frac{1}{i\hbar}[\hat f,\hat H]_- + \frac{1}{2}\lbrace \hat f,\hat
H\rbrace - \frac{1}{2} \lbrace \hat H,\hat f\rbrace = {\cal \hat I}[\hat
f], 
\eeq 
where $[A,B]_{\mp}$ denotes the commutator (resp.
anticommutator) of $A$ and $B$ and $\lbrace A,B
\rbrace=\nabla_{\bfr}A\nabla_{\bfp} B-\nabla_{\bfp}A\nabla_{\bfr} B$
designates the classical Poisson brackets. 
${\cal \hat I}[\hat f]$ represents the spin operator for the collision term.
In the absence of collision, we obtain the following coupled set of equations 
of motion for the particle and spin components of the distribution function:
\beq
\ba{rrl}
(\partial_t +\frac{\bfp}{m}\nabla_{\bfr} +e\bfE \nabla_{\bfp} )f
& -\frac{\hbar \omega_{sd}}{2}\nabla_{\bfr}\bfn\cdot \nabla_{\bfp}\bff&=0,\\
(\partial_t +\frac{\bfp}{m}\nabla_{\bfr} +e\bfE \nabla_{\bfp} )\bff
& -\frac{\hbar \omega_{sd}}{2}\nabla_{\bfr}\bfn \nabla_{\bfp}f &\\
&-\omega_{sd} \bfn \times \bff&={\bf 0}.
\ea 
\eeq 
In these equations the
symbols $\times$ and $\cdot$ mean vector product and scalar product
of spin-vectors. 
For spatial quantities written in bold an
implicit scalar product is understood, namely
$\frac{\bfp}{m}\nabla_{\bfr} \equiv \sum_i
\frac{p_i}{m}\nabla_{r_i}$ and $\nabla_{\bfr}\bfn\cdot
\nabla_{\bfp}\bff \equiv \sum_i \nabla_{r_i}\bfn \cdot
\nabla_{p_i}\bff $.
For later use, we further write $\hat f=\hat f^0+\hat g$ to
separate the "equilibrium"  ($E=0$, but $\bfn$ may depend on $t$) contribution
$\hat f^0$ from the
out-of-equilibrium contribution $\hat g$; and accordingly for each
component.

\subsection{equilibrium properties}

In the case of no applied electric field, the above equations have simple
solutions.

\paragraph{Homogenous ferromagnet: $\bfn(\bfr,t)\equiv \bfn_0$}

In the absence of an electric field and for a time and space independent direction
$\bfn(\bfr,t)=\bfn_0$ the stationary distributions are well known:
\beq \ba{l}
f^0(\bfp)=\frac{1}{2}(n_F(\epsilon_{\bfp}^+)+n_F(\epsilon_{\bfp}^-)),\\
\bff(\bfp)=f^0 _{\parallel}(\bfp)\bfn_0,\\
\textrm{with}\\
f^0 _{\parallel}(\bfp)=\frac{1}{2}(n_F(\epsilon_{\bfp}^+)-n_F(\epsilon_{\bfp}^-)),
\ea
\eeq
where $\epsilon_{\bfp}^{\pm}=\frac{\bfp^2}{2m}\pm \frac{\hbar \omega_{sd}}{2}$
and $n_F(\epsilon)=(e^{\beta(\epsilon-\mu)}+1)^{-1}$ is the Fermi statistic.
The physical quantities are then:
\beq \ba{l}
n(\bfr,t)=n_e=\int \textrm{d}\epsilon \nu(\epsilon) f^0(\epsilon),\\
\bfm(\bfr,t)=m_{\parallel}\bfn_0,\\
\textrm{with}\\
m_{\parallel} = \mu_B P n_e=-\mu_B\int \textrm{d} \epsilon \nu(\epsilon) 
f^0 _{\parallel}(\epsilon),
\ea
\eeq
that defines the polarization $P$.
Note that the scalar functions $f^0$ and $f^0 _{\parallel}$ depend in fact
of the energy $\epsilon$.

\paragraph{Inhomogenous ferromagnet $\bfn(\bfr,t)$}

The naive extension to a space and time dependent $\bfn(\bfr,t)$ would be an unchanged
particle
distribution and a magnetization distribution that follows $\bfn(\bfr,t)$ adiabatically:
$\bff(\bfp,\bfr,t)=f^0 _{\parallel}(\bfp)\bfn(\bfr,t)$.
(Un)fortunately it is not the solution of the equations of motion above.
Instead, to linear order in the space-time gradients, the solution is of the form:
\beq
\ba{ll}
\bff(\bfp,\bfr,t)=&f^0 _{\parallel}(\bfp)\bfn(\bfr,t)\\
&+\ell_{sd} f^0 _{1 \bfr}(\bfp) \bfn \times \nabla_{\bfr} \bfn +
\tau_{sd} f^0 _{1 t}(\bfp) \bfn \times \partial_t \bfn
\ea
\eeq
with
\beq
\ba{l}
f^0 _{1 t}(\bfp)=-f^0_{\parallel}(\bfp),\\
f^0 _{1 \bfr}(\bfp)=-\frac{\tau_{sd}}{\ell_{sd}}\frac{\bfp}{m}f_{\perp}(\epsilon),\\
\ea
\eeq
and
\beq
f_{\perp}(\epsilon)=f^0 _{\parallel}(\epsilon)-
\frac{\hbar \omega_{sd}}{2}\partial_{\epsilon}f^0 (\epsilon).
\label{eq:fperp}
\eeq
The above defined function $f_{\perp}(\epsilon)$ will appear many times
in this paper
and constitutes a key ingredient of most transverses magnetic quantities.
The figure below plots this function $f_{\perp}(\epsilon)$ compared to
$f^0 _{\parallel}(\epsilon)$.
\begin{figure}[h]
\includegraphics[scale=0.5]{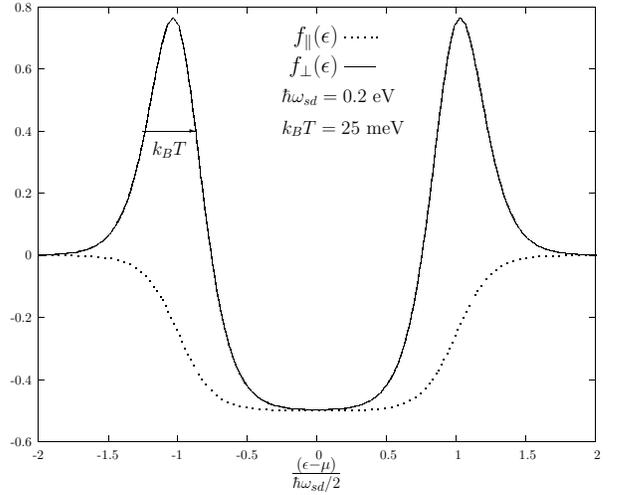}
\caption{The two functions $f_{\perp}(\epsilon)$ and $f^0 _{\parallel}(\epsilon)$
for parameters $k_BT=0.025 meV$ and $\hbar \omega_{sd}=0.2 eV$.}
\end{figure}

The physical consequences of these non adiabatic components are the
following.
The itinerant electrons magnetization is
$\bfm(\bfr,t)=m_{\parallel}\bfn(\bfr,t)+\tau_{sd} m_{1 t}
\bfn(\bfr,t)\times \partial_t \bfn$ with $m_{1 t} =-m_{\parallel}$.
As already pointed out by Zhang and Li \cite{zhangli}, in the $s-d$ picture, the
back action from the component $m_{1 t}$ renormalizes the $\gamma$
term of the $d$ electrons magnetization LLG equation.
The component
$f^0 _{1 \bfr}(\bfp)$ does not contribute to the magnetization
because of its rotational symmetry in momentum space.
However it contributes to a local
equilibrium spin current tensor $\bfJ(\bfr,t)=\ell_{sd} J_{1
\bfr}\bfn(\bfr,t)\times \nabla_{\bfr} \bfn$ with
\beq
\ba{ll}
{J_{1}}_j^{i}
&= -\mu_B \int d\tau \frac{p^i}{m} f^0 _{1 j}(\bfp)\\
&=\delta^{ij}\frac{2\tau_{sd}}{3m\ell_{sd}}\mu_B
\int \textrm{d}\epsilon \nu(\epsilon)\epsilon f_{\perp}(\epsilon)
\ea
\eeq
where we used the identity
$\int \dfrac{\textrm{d}\hat \bfp}{4\pi} \hat \bfp^i \hat \bfp^j=\frac{\delta^{ij}}{3}$.
In the $s-d$ picture, this equilibrium spin current has no back
action.
In the itinerant ferromagnet picture where conduction
electrons are responsible for the local magnetization itself (e.g.
$\Delta_{sd}$ is the exchange interaction between the electrons),
the divergence of this equilibrium spin current should be equivalent
to the local exchange field term, as pointed out by several authors
\cite{Slonczewski89,Stiles06}.
This equilibrium spin current cannot be derived from the ZL macroscopic approach.
As we will show in a forthcoming paper, it is also obtained with the quasiclassical
Keldysh approach.

\section{spin-conserving impurity scattering}

\subsection{Collision integral and transport equations}

We now consider the scattering by impurities in a purely
phenomenological manner. For spin-conserving scattering, a first
guess is a local collision integral of the form
\beq
\ba{lll}
\hat{\cal I}[\hat f(\bfp,\bfr,t)]
&= {\cal I} \hat I+{\mbox{\boldmath ${\cal I}$}} \cdot \hat \bfsigma\\
&=\int d\tau'
w_{\bfp',\bfp}\hat f(\bfp',\bfr,t)(1-\hat f(\bfp,\bfr,t))\\
& \ \  -w_{\bfp,\bfp'}\hat f(\bfp,\bfr,t)(1-\hat f(\bfp',\bfr,t)),
\ea
\eeq
where $w_{\bfp,\bfp'}$ is the probability to scatter from momentum $\bfp$ to
momentum $\bfp'$.
In the expression above $(1-\hat f(\bfp,\bfr,t))$ means
$(1-f)\hat I -\bff \cdot \hat \bfsigma$.
In general $\bff(\bfp',\bfr,t)$ can be non collinear to $\bff(\bfp,\bfr,t)$.
As a consequence, due to the Pauli matrices properties, the above form leads to an
unphysical imaginary contribution like
$i(w_{\bfp',\bfp}+w_{\bfp,\bfp'})(\bff(\bfp',\bfr,t)
\times \bff(\bfp,\bfr,t))\cdot \hat \bfsigma$.
To prevent such a contribution it is necessary to antisymmetrize correctly the
previous expression:
\beq
\ba{rl}
\hat {\cal I}=\frac{1}{2}\int d\tau'
&w_{\bfp',\bfp} \left[\hat f(\bfp',\bfr,t),(1-\hat f(\bfp,\bfr,t))\right]_+\\
-&w_{\bfp,\bfp'}\left[\hat f(\bfp,\bfr,t),(1-\hat f(\bfp',\bfr,t))\right]_+.
\ea
\eeq
With this form, the equations of motion for the charge and spin components of the
distribution function become:
\beq
\ba{rrl}
(\partial_t +\frac{\bfp}{m}\nabla_{\bfr} +e\bfE \nabla_{\bfp} )f
& - \frac{\hbar \omega_{sd}}{2}\nabla_{\bfr}\bfn\cdot \nabla_{\bfp}\bff
&={\cal I},\\
(\partial_t +\frac{\bfp}{m}\nabla_{\bfr} +e\bfE \nabla_{\bfp} )\bff
& -\frac{\hbar \omega_{sd}}{2}\nabla_{\bfr}\bfn\cdot \nabla_{\bfp}f&\\
&- \omega_{sd}\bfn\times \bff&={\mbox{\boldmath ${\cal I}$}},
\ea
\eeq
with ${\cal I}=\int d\tau' w_{\bfp,\bfp'}( f(\bfp')-f(\bfp))$
(similarly for ${\mbox{\boldmath ${\cal I}$}}$) and where we use the
property that for elastic scattering
$w_{\bfp,\bfp'}=w_{\bfp',\bfp}$.

Whatever the form of $w_{\bfp,\bfp'}$, by integrating these equations of motion over $\bfp$
the resulting continuity equations for the particle density $n(\bfr,t)$ and local
magnetization $\bfm(\bfr,t)$ are the same as in the collisionless situation in absence
of electric field, namely:
\beq
\ba{l}
\partial_t n(\bfr,t)+ \textrm{div} j(\bfr,t)=0,\\
\partial_t \bfm(\bfr,t)+\textrm{div} \bfJ(\bfr,t)-
\omega_{sd}\bfn(\bfr,t)\times \bfm(\bfr,t) =0.
\ea
\eeq
As a consequence, the magnetization in the presence of spin-conserving scattering and
DC electric field driving the system out-of-equilibrium
cannot have finite components for
either $m_{2\bfr} \nabla_{\bfr} \bfn$ or $m_{2t} \partial_t \bfn$.
Thus, in that situation, as back action there are no induced $\alpha$ and $\beta$
torque terms.
We stress that this somewhat disappointing result concerning the nonexistence of a
$\beta$ term (finite $m_{2\bfr}$) for macroscopic quantities is not characteristic
of what happens to the  microscopic distribution function.
In fact, as we explain below, at the level of the spin distribution distribution
function, in the presence of spin-conserving impurities in out-of-equilibrium situation,
there is a finite energy resolved magnetization $m_{2\bfr}(\epsilon)$ that results
after angular momentum integration
(or a finite angular momentum resolved magnetic density tensor $m_{2\bfr}(\hat \bfp)$
that results after energy integration).
So, the itinerant electron magnetization is not far from having a finite $m_{2\bfr}$.

Spin-conserving scattering
by impurities produces other interesting results that we now
describe in both equilbrium and out-of-equilibrium situations.
With our general Ansatz
linear in the gradients, we obtain the following set of coupled equations for
each "stationary" component:
\beq
\ba{l}
e\bfE \nabla_{\bfp} f={\cal I},\\
e\bfE\nabla_{\bfp} f_{\parallel}=
{\cal I}_{\parallel},\\
e\bfE  \nabla_{\bfp} f_{2 \bfr}+\frac{1}{\ell_{sd}}(\frac{\bfp}{m}f_{\parallel}+
\frac{\hbar \omega_{sd}}{2}\nabla_{\bfp}f)+\omega_{sd}f_{1 \bfr}={\cal I}_{2 \bfr},\\
e\bfE \nabla_{\bfp} f_{1 \bfr}-\omega_{sd}f_{2 \bfr}=
{\cal I}_{1 \bfr},\\
e\bfE \nabla_{\bfp} f_{2 t}+\frac{1}{\tau_{sd}}f_{\parallel}+\omega_{sd}f_{1 t}=
{\cal I}_{2 t},\\
e\bfE \nabla_{\bfp} f_{1 t}-\omega_{sd}f_{2 t}=
{\cal I}_{1 t},\\
\ea
\eeq
The components $f,f_{\parallel}$ are completely decoupled from the others.
By contrast, $f_{1 \bfr},f_{2 \bfr}$ are coupled and depend on
$f,f_{\parallel}$, and $f_{1 t},f_{2 t}$ are coupled and depend
only on $f_{\parallel}$.

\subsection{equilibrium properties}

To go further we need to specify the form of the
probability $w_{\bfp,\bfp'}$.
To simplify the
calculation we only consider the case of isotropic scattering
$w_{\bfp,\bfp'}=w(\epsilon_{\bfp})\delta(\epsilon_{\bfp}-\epsilon_{\bfp'})$.
As usual we define the inverse scattering time $\tau$ by:
\begin{equation}
\frac{1}{\tau(\epsilon_{\bfp})} \equiv
\int \dfrac{\textrm{d}\bfp'}{(2\pi \hbar)^3}w(\epsilon_{\bfp})
\delta(\epsilon_{\bfp}-\epsilon_{\bfp'})
=w(\epsilon_{\bfp})\nu(\epsilon_{\bfp})
\end{equation}
and $\eta(\epsilon_{\bfp})=\frac{\tau_{sd}}{\tau}$.
The ratio $\eta$ is anticipated to be much smaller than unity.
Nevertheless, the calculations will be performed for a general $\eta$.
With this specific form and in the absence of an electric field the
previous expressions of $f(\bfp),f_{\parallel}(\bfp),f_{1 t}(\bfp)$,
and $f_{2 t}(\bfp)=0$ are still solutions; their corresponding
collision integral vanishes because all these distributions are
function of $\epsilon_{\bfp}$ only. By contrast, the previous
form of $f_{1 \bfr}(\bfp)$ is such that for a general $w_{\hat
\bfp,\hat \bfp'}$ (and in particular for isotropic scattering), the
corresponding collision integral ${\cal I}_{1 \bfr}$ does not
vanish. It thus implies a non vanishing $f_{2 \bfr}(\bfp)$ component
that may modify $f_{1 \bfr}(\bfp)$ as a back action.
One finds that the new self consistent equilibrium solutions
$f^0_{1 \bfr}(\bfp)$, $f^0_{2 \bfr}(\bfp)$ are:
\beq
\ba{l}
f^0_{1 \bfr}(\bfp)=-\frac{\tau_{sd}}{\ell_{sd}}\frac{\bfp}{m}f_{\perp}(\epsilon),\\
f^0_{2 \bfr}(\bfp)=-\eta \frac{\tau_{sd}}{\ell_{sd}} \frac{\bfp}{m}f_{\perp}(\epsilon),\\
\ea
\eeq
where $f_{\perp}$ is as given in (\ref{eq:fperp}) but divided by
$1+\eta^2$.
The main modification induced by the spin-conserving scattering is thus that,
in equilibrium,
the spin current has now two components: a component collinear to
$-\bfn(\bfr,t)\times \nabla_{\bfr} \bfn$ and a component collinear to
$-\nabla_{\bfr} \bfn$, of smaller amplitude by a factor $\eta=\frac{\tau_{sd}}{\tau}$.
The total modulus of the equilibrium spin current is smaller by a factor
$\frac{1}{\sqrt{1+\eta^2}}$ compared to the collisionless situation.
Apart from this change of modulus, the scattering has thus induced a rotation in
spin space of the equilibrium spin current vector.
The ratio between the two transverse components of the current
defines a rotation angle $\theta$
by $\tan{\theta}=\eta=\frac{\tau_{sd}}{\tau}$.
Let us stress that it is rather unusual to find any modification of the equilibrium
distribution by elastic scattering;
it might be a hint that some "quantum" corrections have been neglected.

From now on, to simplify the expressions, we assume that
$\tau(\epsilon)$ is constant and independent of energy.
When necessary, we comment on the validity of our results for an energy dependent
$\tau(\epsilon)$.

\subsection{out-of-equilibrium properties}

Let us now consider the effect of the electric field.
In the presence of an infinitesimal electric field, to each previous component
$f_{\alpha}(\bfp)$
will be added an out-of-equilibrium component
$g_{\alpha}(\bfp)\equiv g_{\alpha}(\hat \bfp,\epsilon)$.
To linear order in electric field and for isotropic scattering, standard calculations
lead to the solutions
\beq
\ba{l}
g(\bfp)=-e\bfE\frac{\bfp}{m}\tau \partial_{\epsilon} f^0(\epsilon),\\
g_{\parallel}(\bfp)=-e\bfE\frac{\bfp}{m}\tau \partial_{\epsilon}
f^0 _{\parallel}(\epsilon).
\ea
\eeq
The components $g_{1 t},g_{2 t}$ are solutions of the following
coupled equations:
\beq \ba{l}
\frac{1}{\tau_{sd}}g_{\parallel}+\omega_{sd}g_{1 t}=
\frac{1}{\tau} \int \dfrac{\textrm{d}\hat \bfp'}{4\pi}(g_{2 t}(\hat \bfp',\epsilon)
-g_{2 t}(\hat \bfp,\epsilon)),\\
e\bfE\nabla_{\bfp}f^0_{1 t}-\omega_{sd}g_{2 t}=
\frac{1}{\tau} \int \dfrac{\textrm{d}\hat \bfp'}{4\pi}( g_{1 t}(\hat \bfp',\epsilon)
-g_{1 t}(\hat \bfp,\epsilon)).
\ea \eeq
The solutions are immediate:
\beq  \ba{l}
g_{1 t}(\bfp)=-g_{\parallel}(\bfp),\\
g_{2 t}(\bfp)=0.
\ea \eeq
The last two components are solutions of the following coupled equations:
\beq \ba{l}
e\bfE\nabla_{\bfp}f^0_{2 \bfr}+\frac{1}{\ell_{sd}}(\frac{p}{m}g_{\parallel}-
\frac{\hbar \omega_{sd}}{2}\nabla_{\bfp}g)+\omega_{sd}g_{1 \bfr}\\
 \ \ \ \ \ \ \ \ \ \ = \frac{1}{\tau} \int
 \dfrac{\textrm{d}\hat \bfp'}{4\pi}(g_{2 \bfr}(\hat \bfp',\epsilon)-
 g_{2 \bfr}(\hat \bfp,\epsilon)),\\
e\bfE\nabla_{\bfp}f^0_{1 \bfr}-\omega_{sd}g_{2 \bfr}\\
 \ \ \ \ \ \ \ \ \ \ =\frac{1}{\tau} \int \dfrac{\textrm{d}\hat \bfp'}{4\pi}(
 g_{1 \bfr}(\hat \bfp',\epsilon)-g_{1 \bfr}(\hat \bfp,\epsilon)).
\ea \eeq
The general solutions of these equations are linear in $E$
and have the parametric form:
\beq \ba{l} g_{1 \bfr}\bfn \times
\nabla_{\bfr} \bfn
\equiv {g_{1}}_i^{j} \hat E^j \bfn \times \nabla_{r_i} \bfn\\
\textrm{with}\\
{g_{1}}_i^{j}=\frac{\tau_{sd}^2}{\ell_{sd}} \frac{e|E|}{m}
[\hat \bfp^i \hat \bfp^j x_1(\epsilon) +\delta^{ij} y_1(\epsilon)]
\ea \eeq
and similarly for $g_{2 \bfr}\nabla_{\bfr}\bfn$.
For the quantities $x_{1,2}(\epsilon)$ and $y_{1,2}(\epsilon)$ we obtain
finally:
\beq \ba{l}
x_1=\frac{1+3\eta^2}{\eta}\frac{2\epsilon}{1+\eta^2}\partial_{\epsilon}f_{\perp}\\
x_2=2\eta^2\frac{2\epsilon}{1+\eta^2} \partial_{\epsilon}f_{\perp}\\
y_1=\frac{\eta}{3}x_2+\frac{1+2\eta^2}{\eta}f_{\perp}-\frac{1}{\eta}f^0_{\parallel}\\
y_2=-\frac{\eta}{3}x_1-f_{\perp}
\ea\eeq
Even if $g_{1 \bfr}$ and $g_{2 \bfr}$ are not simple distributions,
from their rotational symmetry properties in momentum space
we deduce that there is no finite spin current associated.
By contrast, it is not clear if any finite magnetization will survive after angular
and energy integration.
Thus, it appears interesting to define two partial magnetization
densities for each components.
On the one hand, the energy resolved transverse magnetization densities
$m_{1 \bfr}(\epsilon), m_{2\bfr}(\epsilon)$
are obtained after integration over angles alone,
and on the other hand the angle resolved magnetization tensors
$m_{1\bfr}(\hat \bfp), m_{2\bfr}(\hat \bfp)$ are obtained after energy integration.
These quantities are defined through:
\beq
\ba{ll}
m_{1 \bfr}(\bfr,t)&=-\mu_B \int d\tau g_{1 \bfr}(\bfr,\bfp,t)\\
&=\frac{\tau_{sd}^2}{\ell_{sd}} \frac{e|E|}{m} \int   d\epsilon \
m_{1\bfr}(\epsilon),\\
&=\frac{\tau_{sd}^2}{\ell_{sd}} \frac{e|E|}{m} \int \displaystyle{\frac{d\hat \bfp}{4\pi}}
 m_{1\bfr}(\hat \bfp),
\ea \eeq
and similarly for $m_{2 \bfr}(\bfr,t)$.
Performing the angular integration,  we obtain the following expression for the
energy resolved transverse magnetization densities:
\beq
\ba{ll}
m_{1\bfr}(\epsilon)
&=\frac{1+2\eta^2}{\eta}m_{\perp}(\epsilon)-\frac{1}{\eta}m_{\parallel}(\epsilon),\\
m_{2 \bfr}(\epsilon)
&=-m_{\perp}(\epsilon),\\
\textrm{with}&\\
m_{\perp}(\epsilon)&=-\mu_B \nu(\epsilon)(\frac{2}{3}\epsilon \partial_{\epsilon}
f_{\perp}+f_{\perp}),\\
m_{\parallel}(\epsilon)&=-\mu_B \nu(\epsilon)f_{\parallel}^0.\\
\ea
\eeq
The form of $m_{\perp}(\epsilon)$ makes it clear why
$\int d\epsilon \ m_{\perp}(\epsilon)=0$.
A possible very challenging experimental test of our approach would be to
measure, with a magnetic STM tip, these effective itinerant electrons energy resolved
transverse magnetic densities as a function of energy.
In fact the existence of a finite $m_{2\bfr}(\epsilon)$ in the absence of any spin
flip scattering would be a proof of the non trivial component $f_{\perp}$ that is
also at the origin of the equilibrium spin current.

Performing only the energy integration gives rise to the two angle resolved
magnetic tensors $m_{2\bfr}(\hat \bfp)$ and $m_{1\bfr}(\hat \bfp)$:
\beq
\ba{l}
m_{1 \bfr}(\hat \bfp)=\frac{1+3\eta^2}{\eta(1+\eta^2)} m_{\perp}(\delta^{ij}-
3\hat \bfp^i \hat \bfp^j)-\frac{1}{\eta}m_{\parallel} \delta^{ij},\\
m_{2 \bfr}(\hat \bfp)=\frac{2\eta^2}{1+\eta^2} m_{\perp}(\delta^{ij}-3\hat \bfp^i \hat \bfp^j),\\
\textrm{with}\\
m_{\perp}=-\mu_B \int d\epsilon \ \nu(\epsilon)f_{\perp}(\epsilon),\\
m_{\parallel}=-mu_B \int d\epsilon \ \nu(\epsilon)f_{\parallel}(\epsilon),
\ea
\eeq
These last two expressions show why a further angular integration cancels all the
contributions induced by the existence of the non trivial component $f_{\perp}(\epsilon)$.

All these results are, in fact, valid for any spatial dimension provided we use the
corresponding density of states and angular integration.
They can also be generalized to an energy dependent scattering time with more
complicated expressions for the distributions $g_{1,2 \bfr}$.
From the spin continuity equation, we know that we should find
$m_{1 \bfr}=-\frac{\tau_{sd}}{\ell_{sd}} j_{\parallel}$.
This is indeed the case because we have the equality
\beq
\int d\tau {g_{1}}_i^{j}(\bfp)
=-\frac{\tau_{sd}}{\ell_{sd}}\int d\tau
\frac{p^i}{m}{g_{\parallel}}^j(\bfp).
\eeq
For an energy independent scattering time we furthermore obtain:
$m_{1 \bfr}=-\frac{\tau_{sd}}{\ell_{sd}} j_{\parallel}=
-\frac{\tau_{sd}}{\ell_{sd}}\frac{e|E|\tau}{m} m_{\parallel}$.

Note that, although our results are non perturbative in $\eta$, they
are valid order by order in $\eta$ taking care of the fact that the leading order is
$1/\eta$.
This is in contrast with the linear response results of Kohno {\it et al.} where the
standard "leading order term" leads to an unphysical magnetization component
$m_{2 \bfr}$ that is cancelled only when vertex corrections that constitute
infinite order resummation are carefully taken into account.

We can partially generalize our results in two ways \cite{generalisation1}:
(i) define distinct intraband
 scattering probabilities $w^{\pm}_{\bfp,\bfp'}$ for energies $\epsilon_{\pm}$.
(ii) define anisotropic scattering probability $w_{\bfp,\bfp'}$.

\section{Spin-flip impurity Scattering}

\subsection{Spin-basis invariant collision integral}

To extend our Boltzmann approach to spin-flip scattering, a key step is to find a
spin basis invariant formulation of a collision integral that characterizes a
spin-flip process.
We explain in details in the Appendix how to find 
this collision integral in the case of uniform magnet, and how to
phenomenologically generalize it to non uniform magnets.
The resulting collision integral is:
\beq \ba{l} \hat
{\cal I}^{\textrm{sf}}=\frac{1}{4}\int d\tau' \left[\hat
w^{\textrm{sf}}_{\bfp,\bfp'}(\bfr,t),
[\hat{\bar f}(\bfp',\bfr,t),1-\hat f(\bfp,\bfr,t)]_+\right]_+\\
\ \ \ \ \ \ \ \ \ \ \ \ \ \ -\left[\hat w^{\textrm{sf}}_{\bfp,\bfp'}(\bfr,t),
[\hat f(\bfp,\bfr,t),1-\hat{\bar f}(\bfp',\bfr,t)]_+ \right]_+,\\

\textrm{with} \\

\hat w^{\textrm{sf}}_{\bfp,\bfp'}(\bfr,t)=
[\frac{1}{2}(w^{\textrm{sf}+}_{\bfp,\bfp'}+ w^{\textrm{sf}-}_{\bfp,\bfp'})\hat I\\
 \ \ \ \ \ \ \ \ \ \ \ \ \ \ \ \ -\frac{1}{2}
 (w^{\textrm{sf}+}_{\bfp,\bfp'}-w^{\textrm{sf}-}_{\bfp,\bfp'})
\bfn(\bfr,t) \cdot \hat \bfsigma],\\

\textrm{and}\\

\hat{\bar f}(\bfp,\bfr,t)\equiv f(\bfp,\bfr,t) \hat I -\bff_{\parallel}(\bfp,\bfr,t)
\cdot \hat \bfsigma
\ea \eeq
The fact that $\hat{\bar f}(\bfp,\bfr,t)$ appears in the collision integral is quite
natural since there is spin flip. 
On the other hand, the above form
of $\hat {\cal I}^{\textrm{sf}}$ implies an effective vector spin
flip probability $\hat w^{\textrm{sf}}_{\bfp,\bfp'}$ that depends on
the magnetization direction $\bfn(\bfr,t)$.
Note that there is {\it a priori} no reason to prevent contributions to
$\hat w^{\textrm{sf}}_{\bfp,\bfp'}(\bfr,t)$ that are linear in gradients of $\bfn(\bfr,t)$.
But with our phenomenological approach there is no way to guess their specific form.
This is in fact another key reason why a quantum approach using Keldysh Green function
technique might be useful.

In the collision integral shown above, due to the appropriate antisymetrization 
and for elastic scattering, all the terms that are quadratic in the distribution 
functions cancel. 
It then appears that in the equations of motion detailed in the Appendix, the
effective spin flip collision
integrals that appear for each component have the forms:
 ${\cal I}^{\textrm{sf} \pm}=\int d\tau'
\frac{1}{2}(w^{\textrm{sf}+}_{\bfp,\bfp'}\pm w^{\textrm{sf}-}_{\bfp,\bfp'})
(f(\bfp')-f(\bfp))$
and
${\cal I}^{\textrm{sf} \pm}_{\alpha}=\int d\tau'
\frac{1}{2}(w^{\textrm{sf}+}_{\bfp,\bfp'}\pm w^{\textrm{sf}-}_{\bfp,\bfp'})
(f_{\alpha}(\bfp')+f_{\alpha}(\bfp))$ for $\alpha=\parallel,1,2$.

In the following we shall consider only isotropic scattering in $\bfp$ space
($w^{\textrm{sf}}_{\hat \bfp,\hat \bfp'}(\epsilon)\equiv w^{\textrm{sf}}(\epsilon)$).
Accordingly we define:
\beq \ba{l}
\dfrac{1}{\tau^{\textrm{sf}}_{\pm}(\epsilon_{\bfp})}=
\int d\tau' w^{\textrm{sf}\mp}_{\bfp,\bfp'}
=w^{\textrm{sf}}(\epsilon_{\bfp}^{\pm})\nu(\epsilon_{\bfp}\pm \hbar \omega_{sd}),\\
\eta^{\textrm{sf}}_{\pm}(\epsilon)=
\dfrac{\tau_{sd}}{\tau^{\textrm{sf} }_{\pm}(\epsilon)},
\ea \eeq
We further define:
\beq \ba{l}
\dfrac{1}{\tau^{\textrm{sf}}}=\frac{1}{2}(\dfrac{1}{\tau^{\textrm{sf}}_+}+
\dfrac{1}{\tau^{\textrm{sf}}_-}),\\
\dfrac{1}{\bar \tau}=\dfrac{1}{\tau}+\dfrac{1}{\tau^{\textrm{sf}}},\\
p^{\textrm{sf}}=\dfrac{\tau^{\textrm{sf}}_+-\tau^{\textrm{sf}}_-}{\tau^{\textrm{sf}}_-+\tau^{\textrm{sf}}_+}<1,
\ea \eeq
and correspondingly $\eta^{\textrm{sf}},\bar \eta$ with
$\eta^{\textrm{sf}} <\bar \eta$.

\subsection{Zhang-Li relaxation time approximation}

In order to recover equations of motion for the macroscopic quantities that correspond 
to the relaxation time approximation used by ZL, one needs
three further assumptions:

(i) $$
\frac{1}{2}(w^{\textrm{sf}+}_{\bfp,\bfp'}+w^{\textrm{sf}-}_{\bfp,\bfp'})\simeq
\frac{1}{\tau^{\textrm{sf}}}
\frac{1}{\nu(\epsilon_{\bfp})}\delta(\epsilon_{\bfp'}-\epsilon_{\bfp}),$$
and
$\tau^{\textrm{sf}\pm}$ are constants independent of the energy
 $\epsilon_{\bfp}$;

(ii) $$\frac{1}{2}(w^{\textrm{sf}+}_{\bfp,\bfp'}-w^{\textrm{sf}-}_{\bfp,\bfp'})
\simeq 0;$$

(iii) with assumption (i), at equilibrium, the collision integral of $f_{\parallel}$  
is no longer equal to zero at it should be.
Therefore one needs to replace $f_{\parallel}$ by 
$g_{\parallel}=f_{\parallel}-f_{\parallel} ^0$
in the corresponding collision integral.

\subsection{Extended relaxation time approximation}

Because of assumption (ii) the usual relaxation time approximation
ignores some important qualitative physics.
In fact it is
possible to relax (ii) because for the two components
$f,f_{\parallel}$ it is not necessary to make any approximation to
obtain their exact forms in both equilibrium and out-of-equilibrium
situations.

Indeed, by substituting the usual equilibrium form
$f^0,f^0_{\parallel}$, it is easily verified that both 
${\cal I}^{\textrm{sf}\pm}\ne 0$ and 
${\cal I}^{\textrm{sf} \pm}_{\parallel}\ne 0$.
Nevertheless
${\cal I}^{ \textrm{sf}+}+{\cal I}^{\textrm{sf}-}_{\parallel}=0$ and 
${\cal I}^{\textrm{sf} -}+{\cal I}^{\textrm{sf}+}_{\parallel}=0$, 
thus the spin flip scattering does not
modify the equilibrium properties of these two components.

By contrast, as we show in the Appendix in out-of-equilibrium situation, by
relaxing (ii) the contributions $g$ and $g_{\parallel}$ are
qualitatively modified. 
For the perpendicular components, without
approximation (i) it is not possible to extract explicit closed
forms for their distribution functions.
Therefore in the following, as main assumption we assume that (i) is valid 
for the collision integral of the perpendicular components.

In the following we only give the final expressions for the distributions and 
physical quantities in both equilbrum and out of equilbrum situations. 
The detailed steps of the calculations are described in the Appendix.

\subsection{equilibrium properties}

At equilibrium $f^0$ and $f^0_{\parallel}$ are unmodified.
For $f^0_{1 \bfr},f^0_{2 \bfr}$ the previous equilibrium forms stay also valid
but with a modified scattering time $\tau \rightarrow \bar \tau$.
Thus at equilibrium the "\bfp-isotropic" spin flip mechanism does not qualitatively
change the physics of these two contributions.

Essential modifications arise for the last two components $f^0_{1 t},f^0_{2 t}$.
At equilibrium they depend only on $\epsilon_{\bfp}$, therefore the
energy and angular integrations of collision integrals can be performed and the two
components are easily found :
\beq \ba{l}
f^0_{1 t}(\epsilon)=-\dfrac{1}{1+(2\eta^{\textrm{sf}})^2}f^0_{\parallel},\\
f^0_{2 t}(\epsilon)=-\dfrac{2\eta^{\textrm{sf}}}{1+(2\eta^{\textrm{sf}})^2}f^0_{\parallel}
\ea \eeq
For energy independent $\eta^{\textrm{sf}}$ the $\bfp$ integration is immediate and we
obtain two perpendicular components to the "equilibrium" magnetization:
a component $m_{1 t}=\frac{1}{1+(2\eta^{\textrm{sf}})^2}m_{\parallel}$
collinear to $-\bfn \times \partial_t \bfn$ and a component
$m_{2 t}=2\eta^{\textrm{sf}}m_{1 t}$ collinear to $-\partial_t \bfn$.
The modulus of this perpendicular magnetization vector is reduced by a factor
$\dfrac{1}{\sqrt{1+(2\eta^{\textrm{sf}})^2}}$ compared to the collisionless
situation.
Once again, apart from this change of modulus, the spin flip has also induced a
rotation in spin space of this transverse magnetization vector.
Quantitatively, the ratio between the two transverse components of the magnetization
defines a new rotation angle
$\tan{\theta^{\textrm{sf}}}=2\eta^{\textrm{sf}}=\dfrac{2\tau_{sd}}{\tau^{\textrm{sf}}}$.
We have previously pointed out that, as back action on the $d$ electron local magnetization, 
the component $m_{1 t}$ contributes to the effective $\gamma$ term \cite{zhangli}.
The new component
$-\tau_{sd} m_{2 t} \partial_t \bfn$ contributes to the effective Gilbert damping
by a term
\beq \ba{ll}
\alpha_{2t}&=\frac{\gamma \Delta_{sd}}{\mu_B
M_s}\tau_{sd}\dfrac{2\eta^{\textrm{sf}}}{1+(2\eta^{\textrm{sf}})^2} m_{\parallel}\\
&= \dfrac{2\tau_{sd}\tau^{\textrm{sf}}}{{\tau^{\textrm{sf}}}^2+(2\tau_{sd})^2}  \frac{\gamma \hbar P n_e}{M_s}
\ea \eeq
This expression exactly coincides with the ZL result
that was calculated using the macroscopic equations of motion in the relaxation 
time approximation. 
Taking only the leading order term in $\tau_{sd}/\tau^{\textrm{sf}}$ we also recover
the expression of Kohno {\it et al} when the appropriate changes of notations are made.

\subsection{out-of-equilibrium properties}

To improve readability,
the transport equations and the explicit calculation
of the out equilibrium distributions
of the different components are described in the Appendix.
The resulting physical properties are the following.

For the particle and parallel spin currents, to first order in $p^{\textrm{sf}}$
we obtain :
\beq \ba{l}
j=\frac{n_e \bar \tau e\bfE}{m}(1+p^{\textrm{sf}}\dfrac{\bar \tau}{\tau^{\textrm{sf}}}P),\\
J_{\parallel}=\mu_B \frac{n_e \bar \tau e\bfE}{m}
(P+p^{\textrm{sf}}\dfrac{\bar \tau}{\tau^{\textrm{sf}}}).
\ea \eeq
Note that, even if there is no difficulty to obtain the results to any order in
$p^{\textrm{sf}}$, the leading order term contains all the important
qualitative modifications.
These expressions show that the spin flip scattering modifies both the particle
current and parallel current in a distinct manner.
Such a result cannot be obtained using the standard relaxation time
approximation and can only be introduced phenomenologically
within the ZL macroscopic approach.
We remark that, when $P=1$ (fully polarized case), the parallel and particle
current coincide as they should.
When $P=0$ (unpolarized) the parallel component is zero because
$p^{\textrm{sf}}$ is implicitly proportional to $P$ (see the definition above).
For later use, we also note that to first order in $p^{\textrm{sf}}$ the relation between
the parallel spin current and the charge current $j_e=ej$ is:
\beq
J_{\parallel}\simeq
\frac{\mu_B}{e}(P+p^{\textrm{sf}}\dfrac{\bar \tau}{\tau^{\textrm{sf}}}(1-P^2)) j_e
\eeq
As explained in the Appendix, in out-of-equilibrium situation the spin current has
also two transverse components collinear to $\partial_t \bfn$ and
$\bfn \times \partial_t \bfn$, respectively.
These two transverse components of the spin current are easily accessible with
our formalism but cannot be explicitly found with the ZL approach and need
to be calculated within the linear response or Keldysh approach.

The last two components $g_{1,2 \bfr}(\bfp)$ of the out-of-equilibrium spin distribution
function, as explained in the Appendix, can be recast in tensor form as
 ${g_{1,2}}_i ^{j}=\frac{\tau_{sd}^2}{\ell_{sd}} \frac{e|E|}{m}[
\hat p^i \hat p^j x_{1,2}(\epsilon)+
\delta^{ij}y_{1,2}(\epsilon)]$, with modified functions $x_{1,2}(\epsilon)$ and
$y_{1,2}(\epsilon)$ as compared to the spin-conserving scattering case.
With these expressions, by performing either angular integration or energy integration,
we can calculate the modified energy resolved magnetic densities
$m_{1 \bfr}(\epsilon), m_{2\bfr}(\epsilon)$ (see Appendix)
and angular resolved magnetic tensors $m_{1 \bfr}(\hat \bfp),m_{2\bfr}(\hat \bfp)$
respectively, and then by further integration the resulting physical magnetization
components $m_{1 \bfr},m_{2\bfr}$.
For comparison, we obtain the following expressions for the modified angular resolved magnetic tensors
$m_{1 \bfr}(\hat \bfp),m_{2\bfr}(\hat \bfp)$ :
\beq
\ba{ll}
m_{1 \bfr}(\hat \bfp)=&
\left(\frac{1+3\bar \eta^2}{\bar \eta(1+{\bar \eta}^2)} m_{\perp}+m_{\perp}^{\textrm{sf}}\right)
(\delta^{ij}-3\hat \bfp^i \hat \bfp^j)\\
&-\frac{1}{\bar \eta} \frac{1}{1+(2\eta^{\textrm{sf}})^2}
\left(m_{\parallel}+p^{\textrm{sf}} \frac{\eta^{\textrm{sf}}}{\bar \eta} n_e \right)\delta^{ij},\\
m_{2 \bfr}(\hat \bfp)=&
\left(\frac{1+2{\bar \eta}^2}{(1+{\bar \eta}^2)} m_{\perp}+\bar \eta m_{\perp}^{\textrm{sf}}\right)
(\delta^{ij}-3\hat \bfp^i \hat \bfp^j)\\
&-\frac{1}{\bar \eta}\frac{2\eta^{\textrm{sf}}}{1+(2\eta^{\textrm{sf}})^2}
\left(m_{\parallel}+p^{\textrm{sf}} \frac{\eta^{\textrm{sf}}}{\eta} n_e \right)\delta^{ij},\\
\textrm{with}&\\
m_{\perp}^{\textrm{sf}}=&-\mu_B \int d\epsilon \ \nu(\epsilon) f_{\perp}^{\textrm{sf}}(\epsilon),\\
f_{\perp}^{\textrm{sf}}(\epsilon)=&p^{\textrm{sf}}\dfrac{\eta^{\textrm{sf}}}{\bar \eta}
\frac{1}{1+{\bar \eta}^2} \frac{1}{\bar \eta}(f^0-\frac{\hbar \omega_{sd}}{2}\partial_{\epsilon} 
f^{0}_{\parallel}),
\ea
\eeq
Performing the angular integration we finally obtain the two transverse magnetization
components:
\beq \ba{ll}
m_{1 \bfr}&=\dfrac{1}{1+(2\eta^{\textrm{sf}})^2}\frac{\tau_{sd} }{\ell_{sd}}J_{\parallel},\\
m_{2
\bfr}&=\dfrac{2\eta^{\textrm{sf}}}{1+(2\eta^{\textrm{sf}})^2}\frac{\tau_{sd}
}{\ell_{sd}}J_{\parallel}. \ea \eeq
As back action on the localized $d$ electron
magnetization, the component $m_{1 \bfr}$ gives rise to the first
spin torque term and thus determines the parameter $u$ in terms of
the charge current density $j_e$:
\beq \ba{ll} u&= \frac{\gamma
\Delta_{sd}}{\mu_B M_s} \ell_{sd}m_{1 \bfr}\\
&=\dfrac{{\tau^{\textrm{sf}}}^2}{{\tau^{\textrm{sf}}}^2+(2\tau_{sd})^2}
(1+p^{\textrm{sf}}\dfrac{\bar
\tau}{\tau^{\textrm{sf}}}\frac{1-P^2}{P})
 \frac{\gamma \hbar P j_e}{e M_s}.
\ea \eeq
The component $m_{2 \bfr}$ gives rise to the second spin
torque term and thus determines the parameter $\beta$ as
\beq
\beta=2\frac{\tau_{sd}}{\tau^{\textrm{sf}}}.
\eeq
This value of $\beta$ coincides with ZL result (it therefore also coincides with
the Kohno result for spin-isotropic spin flip when appropriate modifications of
notations are made).
Note that our result for the value of $u$ cannot be explicitly calculated with
the ZL macroscopic equations of motion.
To leading order in $\frac{\tau_{sd}}{\tau^{\textrm{sf}}}$, the ratio
$\beta/\alpha_{2t} \simeq M_s/m_{\parallel}\ge 1$
and thus, at first sight, it seems compatible with micromagnetics and experiments.
We note however that other sources of dissipation (spin lattice relaxation for example)
might give some contribution to the total effective Gilbert damping parameter
$\alpha$ and thus modify the $\beta/\alpha$ ratio.
It would therefore be interesting if experiments could provide
measurements of the different contributions to the Gilbert damping parameter $\alpha$.




\section{Summary, discussion and perspectives}

Using the Landau-Sillin approach, we have studied the transport of electrons in the
presence of an effective Zeeman field that has a space-time varying direction.
The key ingredient is our Ansatz form of the spin density matrix that consists in
a linear decomposition on quasistationary distribution functions along each possible
direction provided by the first order space-time gradients of the magnetic field direction.
We have shown step by step how the form of the different components of the distribution
function is affected by the presence of spin-conserving and spin flip scattering, in
both equilibrium and out-of-equilibrium situations.
For spin-flip scattering we have defined a spin-basis-invariant collision integral and
an extended relaxation time approximation that show the mixing of the particle and
parallel components.
Our calculations also illustrate the striking difference between a macroscopic
quantity such as the transverse (perpendicular) magnetization component and its 
underlying distribution.
This is particularly clear for the components $m_{2 \bfr}$ and $f_{2 \bfr}(\bfp)$ in
the presence of spin-conserving scattering in the out-of-equilibrium situation.
The term $f_{2 \bfr}(\bfp)$ is non zero and leads to a finite energy resolved magnetization density
$m_{2\bfr}(\epsilon)$ when only an
angular integration $\hat \bfp$ is performed.
But further energy integration of this density gives a zero contribution
as expected from the spin conservation rule.
More qualitatively and physically, we have clearly explained the existence of an
equilibrium spin tranverse current in the direction $\bfn \times \nabla_{\bfr} \bfn$
when there is no scattering at all.
We have shown that, within the Boltzmann approach, the modulus of this equilibrium spin
transverse current and its direction are affected by spin-conserving
and spin flip scattering.
The rotation is by the angle $\bar \theta$  and in the direction $\nabla_{\bfr} \bfn$, compared
to the collisionless situation.
Although we have not calculated it explicitly, we have shown that the out-of-equilibrium
contribution to the transverse spin current is purely in the direction provided by the
time derivative $\bfn\times \partial_t \bfn$ in the spin-conserving situtation.
In presence of spin flip this contribution is further rotated in a complicated
manner towards the direction $\partial_t \bfn$.
Concerning the transverse magnetization components, the situation is somewhat reversed.
In "equilbrum" the transverse magnetization is along $\bfn\times \partial_t \bfn$ in
both collisionless and spin-conserving scattering situations.
In presence of spin flip it is rotated by an angle $\theta^{\textrm{sf}}$ to
the direction $\partial_t \bfn$.
For the localized $d$ electron magnetization this rotation leads to the appearance
of an effective Gilbert damping correction term $\alpha_{2t}$.
In out of equilbrum situation, for spin-conserving scattering the transverse magnetization
is along $\bfn \times \nabla_{\bfr} \bfn$ only and its modulus is directly proportional
to the parallel spin current.
In the presence of spin flip, this transverse magnetization is rotated by an angle
$\theta^{\textrm{sf}}$ in the direction $\nabla_{\bfr} \bfn$.
For the localized $d$ electron magnetization these two components lead to the two spin
transfer torque terms and thus allow determining the two parameters $\beta$ and $u$.
Our results for $\beta$ and $\alpha_{2t}$ exactly with those of the ZL macroscopic approach
and of Kohno {\it et al.} in the spin-isotropic spin flip scattering situation that we consider.
Our approach allows to calculate explicitly the spin current polarization $P_j=u/u_e$
where $u_e$ is the electron drift velocity that is purely phenomenological in ZL
macroscopic approach.


We believe that this paper can be extended in at least four directions
\cite{inpreparation}.

(i)
For spin flip scattering, as we already pointed out, there is no reason to prevent
terms proportional to the gradient of $\bfn(\bfr,t)$ in
the spinor probability $\hat w_{\bfp,\bfp'}(\bfr,t)$.
A natural extension of our work would consist in considering such terms by assuming
some specific forms for the corresponding probability
$\hat w_{\alpha,\bfp,\bfp'}$ ($\alpha=1t,1\bfr,2t, 2\bfr$).
In fact we believe that $\hat w_{1,2 \bfr,\bfp,\bfp'}$ could be at the origin of
an out-of-equilibrium spin Hall current.

(ii) In our paper we have only considered the most simple quadratic dispersion relation
$\epsilon_{\bfp}=\frac{\bfp^2}{2m}$ and a $\bfp$ independant effective Zeeman field.
In the spirit of the work of J. Zhang {\it et al.} \cite{antropov}, it would be
interesting
to explore how far the equilibrium and out of equilbrum properties are changed for a
general dispersion relation and a $\bfp$ dependent effective Zeeman field.

(iii)
It is possible to adapt our method to spin-valve systems \cite{spinvalve}.
The key point consists in expanding the spinor distribution function
$\hat f^{\ell}(\bfp,\bfr,t)$ in each region $\ell=L,R,C$ ($L\equiv$ left thick magnetic
layer, $C\equiv$ central non magnetic layer and $R\equiv$ right thin magnetic layer)
into the most general basis to first
order in time gradients of the two magnetic layers directions $\bfn_{L,R}(t)$.
If for simplicity one assumes that the "thick layer" direction $\bfn_L$
is time independent, naively our Ansatz distribution for the thin layer would be:
\beq \ba{ll}
\hat f^{R}(\bfp,\bfr,t)=&f^{R}(\bfp,\bfr) \hat I+\bff^{R}(\bfp,\bfr,t) \cdot \bfsigma,\\
\textrm{with}&\\
\bff^{R}(\bfp,\bfr,t)=&f_{\parallel}^{R}(\bfp,\bfr)\bfn_{R}(t)\\
&+ (f_{1\bfr}^{R}(\bfp,\bfr)\bfn_{R}\times \bfn_{L}+f_{2\bfr}^{R}(\bfp,\bfr)  \bfn_{L})\\
&+\tau^{R}_{sd} (f_{1t}^{R}(\bfp,\bfr)\bfn_{R}\times \partial_t \bfn_{R}+
f_{2t}^{R}(\bfp,\bfr) \partial_t \bfn_{R})\\
&+\tau^{R}_{sd} f_{3 t}^{R}(\bfp,\bfr)\bfn_{L}\times \partial_t \bfn_{R}
\ea \eeq
In fact, preliminary results show that this Ansatz with an extended basis of six vectors
is still insufficient \cite{remarks1}.

(iv)
In the context of ferromagnetic Fermi liquids, the Landau-Sillin approach has been used
for a long time and equations of motion of the magnetization with terms similar to
the spin torque terms have been established for example by Leggett \cite{leggett}.
Nevertheless, there are still many questions that concern the transverse properties.
To our knowledge, most of the parametrizations used to study these systems
\cite{mineev} are similar to
that of Kohno {\it et al.} and Tserkovnyak {\it et al.}.
We thus believe
that new insights can be provided by a parametrization similar to ours.

As a final remark and anticipating on our paper using Keldysh Green functions,
we have pointed
at several places that a collision integral is intrinsically a
quantum object and that, therefore, quantum corrections might affect the
results.
The microscopic derivation of a collision integral requires
the calculation of a self energy which itself depends on the Green
function.
The quantum object that plays a role
similar to the spinor distribution $\hat f(\bfr,\bfp,t)$ is the
time-space Wigner transform spinor Green function $\hat
G(\bfp,\omega,\bfr,t)$.
Very similarly to the spinor distribution
$\hat f(\bfr,\bfp,t)$ we thus propose to adopt the following
Ansatz form for the Green function:
\beq \ba{ll}
\hat G(\bfp,\omega,\bfr,t)=
&G(\bfp,\omega)\hat I+ {\bf G}(\bfp,\omega,\bfr,t)\cdot \hat \bfsigma\\
\textrm{with}\\
{\bf G}(\bfp,\omega,\bfr,t)=&G_{\parallel}(\bfp,\omega)\bfn(\bfr,t)\\
&+\ell_{sd}\left[G_{1 \bfr}(\bfp,\omega)\bfn\times \nabla_{\bfr}\bfn+
G_{2 \bfr}(\bfp,\omega)\nabla_{\bfr}\bfn\right]\\
&+\tau_{sd}\left[G_{1 t}(\bfp,\omega)\bfn\times \partial_t\bfn+
G_{2 t}(\bfp,\omega)\partial_t \bfn\right].
\ea \eeq

\section{Appendix}

\subsection{Spin-basis invariant collision integral for spin-flip scattering}

Spin flip scattering in a uniform ferromagnet
($\bfn(\bfr,t)\equiv \bfn_0$) corresponds to an "interband" process.
The collision
integral of each eigen-spin distribution function
$f_{\pm}(\bfp)=f(\bfp)\pm f_{\parallel}(\bfp)$ is then:
\beq
\ba{lll} {\cal I}^{\textrm{sf}}_{\pm}&= \int d\tau'  w^{\textrm{sf}
\mp}_{\bfp,\bfp'}&[
f_{\mp}(\epsilon_{\bfp'})(1-f_{\pm}(\epsilon_{\bfp}))\\
&&-f_{\pm}(\epsilon_{\bfp})(1-f_{\mp}(\epsilon_{\bfp'}))]\\
&=\int d\tau' w^{\textrm{sf} \mp}_{\bfp,\bfp'} &
(f_{\mp}(\epsilon_{\bfp'})-f_{\pm}(\epsilon_{\bfp})), \ea \eeq
with $w^{\textrm{sf} \mp}_{\bfp,\bfp'}= w^{\textrm{sf}}_{\hat \bfp,
\hat \bfp'}(\epsilon^{\mp}_{\bfp'})
\delta(\epsilon^{\mp}_{\bfp'}-\epsilon^{\pm}_{\bfp})$.
The corresponding collision integrals for the particle density and
parallel magnetization components are:
\beq \ba{ll} {\cal
I}^{\textrm{sf}}=\int d\tau'
&[\frac{1}{2}(w^{\textrm{sf}+}_{\bfp,\bfp'}+w^{\textrm{sf}-}_{\bfp,\bfp'})
(f(\bfp') -f(\bfp))\\
&+\frac{1}{2}(w^{\textrm{sf}+}_{\bfp,\bfp'}-w^{\textrm{sf}-}_{\bfp,\bfp'})
(f_{\parallel}(\bfp') +f_{\parallel}(\bfp))],\\
{\cal I}^{\textrm{sf}}_{\parallel}=\int d\tau'
&-[\frac{1}{2}(w^{\textrm{sf}+}_{\bfp,\bfp'}-w^{\textrm{sf}-}_{\bfp,\bfp'})
(f(\bfp')-f(\bfp))\\
&-\frac{1}{2}(w^{\textrm{sf}+}_{\bfp,\bfp'}+w^{\textrm{sf}-}_{\bfp,\bfp'})
(f_{\parallel}(\bfp')+f_{\parallel}(\bfp))].
\ea \eeq
To extend the above results to a non uniform magnet we need first to
find a spin-basis-invariant formulation such that for the uniform
ferromagnet we can write a spin-matrix collision integral
$\hat {\cal I}={\cal I}\hat I +{\cal \bf I}\cdot \hat \bfsigma$ directly
in terms of a spin-matrix distribution function
$\hat f(\bfp)=f\hat I+\bff \cdot \hat \bfsigma$
(with $\bff=f_{\parallel}\bfn_0$ for the
uniform case).
The following form of ${\cal I}$ appears to be
compatible with the above results for ${\cal I}$ and
${\cal I}_{\parallel}$:
\beq \ba{l}
\hat {\cal I}^{\textrm{sf}}=\int d\tau'
\hat w^{\textrm{sf}}_{\bfp,\bfp'}
[\hat{\bar f}(\bfp')(1-\hat f(\bfp)) -\hat f(\bfp)(1-\hat{\bar f}(\bfp'))],\\
\textrm{with}\\
\hat w^{\textrm{sf}}_{\bfp,\bfp'}=
[\frac{1}{2}(w^{\textrm{sf}+}_{\bfp,\bfp'}+w^{\textrm{sf}-}_{\bfp,\bfp'})\hat I -
\frac{1}{2}(w^{\textrm{sf}+}_{\bfp,\bfp'}-w^{\textrm{sf}-}_{\bfp,\bfp'})
\bfn_0 \cdot \hat \bfsigma],\\
\textrm{and}\\
\hat{\bar f}(\bfp)\equiv f(\bfp) \hat I -f_{\parallel}(\bfp)\bfn_0
\cdot \hat \bfsigma.
\ea \eeq
On the one hand, the fact that
$\hat{\bar f}(\bfp)$ appears in the collision integral is quite
natural since there is spin flip, on the other hand the above form
of $\hat {\cal I}^{\textrm{sf}}$ implies an effective vector spin
flip probability $\hat w^{\textrm{sf}}_{\bfp,\bfp'}$ that depends on
the magnetization direction.
We now phenomenologically extend the
above form to space-time dependent magnetization with appropriate
antisymmetrization to prevent imaginary terms:
\beq \ba{l} \hat
{\cal I}^{\textrm{sf}}=\frac{1}{4}\int d\tau' \left[\hat
w^{\textrm{sf}}_{\bfp,\bfp'}(\bfr,t),
[\hat{\bar f}(\bfp',\bfr,t),1-\hat f(\bfp,\bfr,t)]_+\right]_+\\
\ \ \ \ \ \ \ \ \ \ \ \ \ \ -\left[\hat w^{\textrm{sf}}_{\bfp,\bfp'}(\bfr,t),
[\hat f(\bfp,\bfr,t),1-\hat{\bar f}(\bfp',\bfr,t)]_+ \right]_+,\\

\textrm{with} \\

\hat w^{\textrm{sf}}_{\bfp,\bfp'}(\bfr,t)=
[\frac{1}{2}(w^{\textrm{sf}+}_{\bfp,\bfp'}+ w^{\textrm{sf}-}_{\bfp,\bfp'})\hat I\\
 \ \ \ \ \ \ \ \ \ \ \ \ \ \ \ \ -\frac{1}{2}(w^{\textrm{sf}+}_{\bfp,\bfp'}-w^{\textrm{sf}-}_{\bfp,\bfp'})
\bfn(\bfr,t) \cdot \hat \bfsigma].
\ea \eeq

\subsection{Transport equation and out of equilbrum distribution functions in the
presence of spin flip scattering}

Using the form
of the collision integrals introduced in the previous section,
the equations of motion of all components in the presence of both spin-flip and
spin-conserving impurities now read:
\beq \ba{l}
e\bfE \nabla_{\bfp} f={\cal I} +{\cal I}^{\textrm{sf}+}+{\cal I}^{\textrm{sf} -}_{\parallel},\\
e\bfE\nabla_{\bfp} f_{\parallel}=
{\cal I}_{\parallel}-{\cal I}^{ \textrm{sf}-}-{\cal I}^{\textrm{sf}+}_{\parallel},\\
e\bfE \nabla_{\bfp} f_{2 \bfr  }+\frac{1}{\ell_{sd}}(\frac{\bfp}{m}f_{\parallel}-
\frac{\hbar \omega_{sd}}{2}\nabla_{\bfp}f)+\omega_{sd}f_{1 \bfr }
={\cal I}_{2 \bfr}-{\cal I}^{\textrm{sf}+}_{2 \bfr},\\
e\bfE \nabla_{\bfp} f_{1 \bfr}-\omega_{sd}f_{2 \bfr}=
{\cal I}_{1 \bfr}-{\cal I}^{\textrm{sf}+}_{1 \bfr},\\
e\bfE \nabla_{\bfp} f_{2 t}+\frac{1}{\tau_{sd}}f_{\parallel}+\omega_{sd}f_{\perp  t 1}=
{\cal I}_{2 t}-{\cal I}^{\textrm{sf}+}_{2 t},\\
e\bfE \nabla_{\bfp} f_{1 t}-\omega_{sd}f_{2 t}=
{\cal I}_{1 t}-{\cal I}^{\textrm{sf}+}_{1 t},\\
\ea \eeq with: ${\cal I}^{\textrm{sf} \pm}=\int d\tau'
\frac{1}{2}(w^{\textrm{sf}+}_{\bfp,\bfp'}\pm w^{\textrm{sf}-}_{\bfp,\bfp'})
(f(\bfp')-f(\bfp))$
and
${\cal I}^{\textrm{sf} \pm}_{\alpha}=\int d\tau'
\frac{1}{2}(w^{\textrm{sf}+}_{\bfp,\bfp'}\pm w^{\textrm{sf}-}_{\bfp,\bfp'})
(f_{\alpha}(\bfp')+f_{\alpha}(\bfp))$ for $\alpha=\parallel,1,2$.

The equilibrium properties are easily found and were described in the main text.
For the out of equilbrum properties, within our extended relaxation time approximation
the equations of motion become:
\beq \ba{l}
e\bfE \nabla_{\bfp} f^0=
-\dfrac{1}{\bar \tau}(g-p^{\textrm{sf}}\dfrac{\bar \tau}{\tau^{\textrm{sf}}}g_{\parallel})+
\dfrac{1}{\bar \tau}\int \frac{d\hat p'}{4\pi}
(g(\bfp')+p^{\textrm{sf}}\dfrac{\bar \tau}{\tau^{\textrm{sf}}}g_{\parallel}(\bfp'))\\

e\bfE \nabla_{\bfp} f_{\parallel}^0=-\dfrac{1}{\bar \tau}(g_{\parallel}-
p^{\textrm{sf}}\dfrac{\bar \tau}{\tau^{\textrm{sf}}}g)\\
\ \ \ \ \ \ \ \ \ \ \ \ \ \ \ \ +\dfrac{1}{\bar \tau}\int \frac{d\hat p'}{4\pi}
(1-2\dfrac{\bar \tau}{\tau^{\textrm{sf}}})g(\bfp')-
p^{\textrm{sf}}\dfrac{\bar \tau}{\tau^{\textrm{sf}}}g(\bfp'))\\

e\bfE \nabla_{\bfp} f^0_{2 t}+\dfrac{1}{\tau_{sd}}g_{\parallel}+\omega_{sd}g_{\perp  t 1}
=-\dfrac{1}{\bar \tau}g_{2 t }\\
\ \ \ \ \ \ \ \ \ \ \ \ \ \ \ \ \ \ \ \ \ \ \ \ \ \ \ \ \ \ \ \ \ \ \ \ \ \ \ \ \
+\dfrac{1}{\bar \tau} \int \frac{d\hat p'}{4\pi}(1-
2\dfrac{\bar \tau}{\tau^{\textrm{sf}}})g_{2 t}(\bfp')\\
e\bfE \nabla_{\bfp} f^0_{1 t}-\omega_{sd}g_{2 t}=-\dfrac{1}{\bar \tau}g_{1 t }
+\dfrac{1}{\bar \tau} \int \frac{d\hat p'}{4\pi}(1-
2\dfrac{\bar \tau}{\tau^{\textrm{sf}}})g_{1 t}(\bfp')\\
e\bfE \nabla_{\bfp} f^0_{2 \bfr  }+\dfrac{1}{\ell_{sd}}(\frac{\bfp}{m}g_{\parallel}-
\frac{\hbar \omega_{sd}}{2}\nabla_{\bfp}g)+\omega_{sd}g_{1 \bfr }
=-\dfrac{1}{\bar \tau}g_{2\bfr }\\
 \ \ \ \ \ \ \ \ \ \ \ \ \ \ \ \ \ \ \ \ \ \ \ \ \ \ \ \ \  \ \ \ \ \ \ \ \ \ \ \ \
 +\dfrac{1}{\bar \tau} \int \frac{d\hat p'}{4\pi}(1-
 2\dfrac{\bar \tau}{\tau^{\textrm{sf}}})g_{2 \bfr}(\bfp')\\
e\bfE \nabla_{\bfp} f^0_{1 \bfr}-\omega_{sd}g_{2 \bfr}
=-\dfrac{1}{\bar \tau}g_{1 \bfr }
+\dfrac{1}{\bar \tau} \int \frac{d\hat p'}{4\pi}(1-
2\dfrac{\bar \tau}{\tau^{\textrm{sf}}})g_{1 \bfr}(\bfp')
\ea
\eeq
The first two equations show that in this extended relaxation time approximation
$g$ and $g_{\parallel}$ are both linear combinations of $\nabla_{\bfp}f^0$ and
$\nabla_{\bfp}f^0 _{\parallel}$.
As a consequence in the first four equations the terms with angular integration over
$\hat \bfp'$ do not contribute.
To solve these equations it is convenient to further split each $g_{\alpha}$ into
$g_{\alpha}=\bar g_{\alpha} +g_{\alpha}^{\textrm{sf}}$ where $\bar g_{\alpha}$ is
obtained, from the expressions found in the spin conserving case, by the substitution
$\tau \rightarrow \bar \tau$.

To simplify the expressions and since it does not qualitatively modify the results,
we only retain the first order contribution in $p^{\textrm{sf}}$ although there is no
difficulty to find the exact forms.
For the first four components, we obtain:
\beq \ba{l}
g^{\textrm{sf}}(\bfp)=
-p^{\textrm{sf}}\dfrac{\eta^{\textrm{sf}}}{\bar \eta}\bar
\tau e\bfE\nabla_{\bfp}f^0 _{\parallel},\\
g^{\textrm{sf}}_{\parallel}(\bfp)=
-p^{\textrm{sf}}\dfrac{\eta^{\textrm{sf}}}{\bar \eta}\bar
\tau e\bfE \nabla_{\bfp} f^0,\\
g^{\textrm{sf}}_{1 t}(\bfp)=
\frac{1}{1+\bar \eta^2}\bar \tau e\bfE
\left[ p^{\textrm{sf}}\dfrac{\eta^{\textrm{sf}}}{\bar \eta}\nabla_{\bfp} f^0
+\dfrac{2\eta^{\textrm{sf}}}{1+(2\eta^{\textrm{sf}})^2}
(1-\bar \eta)\bar \eta\nabla_{\bfp}f^0_{\parallel}\right],\\
g^{\textrm{sf}}_{2 t}(\bfp)=
\frac{\bar \eta}{1+\bar \eta^2}\bar \tau e\bfE
\left[ p^{\textrm{sf}}\dfrac{\eta^{\textrm{sf}}}{\bar \eta}\nabla_{\bfp} f^0
+\dfrac{2\eta^{\textrm{sf}}}{1+(2\eta^{\textrm{sf}})^2}
(1+\bar \eta)\bar \eta\nabla_{\bfp}f^0_{\parallel}\right]
\ea \eeq
Each of these components gives rise to some finite current only (particle or
spin current).

For the last two components $g_{1,2 \bfr}$ we remind that the terms with angular
integration over $\hat \bfp'$ contribute.
Once again we write
 ${g_{1,2}}_i^{j}=\frac{\tau_{sd}^2}{\ell_{sd}} \frac{e|E|}{m}[
\hat p^i \hat p^j x_{1,2}(\epsilon)+
\delta^{ij}y_{1,2}(\epsilon)]$ and split accordingly $x_{1,2}$, $y_{1,2}$ into
$x_{1,2}=\bar x_{1,2} +x_{1,2}^{\textrm{sf}}$ and
$y_{1,2}=\bar y_{1,2} +y_{1,2}^{\textrm{sf}}$.
We then obtain:
\beq \ba{ll}
x^{\textrm{sf}}_{1}=&2\epsilon \partial_{\epsilon}f^{\textrm{sf}}_{\perp},\\
x^{\textrm{sf}}_{2}=&\bar \eta x^{\textrm{sf}}_{1},\\
y^{\textrm{sf}}_{1}=
&-\frac{x^{\textrm{sf}}_{1}}{3}-\dfrac{1}{1+(2\eta^{\textrm{sf}})^2}
\left[2\eta^{\textrm{sf}}(\bar y_2+\frac{\bar x_2}{3})+
(2\eta^{\textrm{sf}})^2(\bar y_1+\frac{\bar x_1}{3}) \right. \\
& \ \ \ \ \ \ \ \ \ \ \ \ \ \ \ \ \ \ \ \ \ \ \ \left. -(1+{\bar \eta}^2)
(\frac{x^{\textrm{sf}}_{1}}{3}+f^{\textrm{sf}}_{\perp})
+p^{\textrm{sf}}\dfrac{\eta^{\textrm{sf}}}{\bar \eta}
\frac{1}{\bar \eta}f^0 \right],\\
y^{\textrm{sf}}_{2}=
&-\bar \eta \frac{x^{\textrm{sf}}_{1}}{3}-\dfrac{2\eta^{\textrm{sf}}}
{1+(2\eta^{\textrm{sf}})^2}
\left[2\eta^{\textrm{sf}}(\bar y_2+\frac{\bar x_2}{3})\right. \\
&\ \ \ \ \ \ \left. -(\bar y_1+\frac{\bar x_1}{3})-
(1+{\bar \eta}^2)(\frac{x^{\textrm{sf}}_{1}}{3}+f^{\textrm{sf}}_{\perp})
+p^{\textrm{sf}}\dfrac{\eta^{\textrm{sf}}}{\bar \eta}\frac{1}{\bar \eta}f^0\right],\\
\textrm{with}&\\
f^{\textrm{sf}}_{\perp}=
&p^{\textrm{sf}}\dfrac{\eta^{\textrm{sf}}}{\bar \eta}\frac{1}{1+{\bar \eta}^2}
\frac{1}{\bar \eta}(f^0-\frac{\hbar \omega_{sd}}{2}\partial_{\epsilon} f^{0}_{\parallel}),
\ea \eeq
With these expressions, by performing the angular integration we obtain the following
modified expressions for the energy resolved
transverses magnetization densities $m_{1\bfr}(\epsilon), m_{2\bfr}(\epsilon)$:
\beq
\ba{ll}
m_{1\bfr}(\epsilon)
&=\frac{1}{1+(2\eta^{\textrm{sf}})^2}
\left(\frac{1+2{\bar \eta}^2+2\eta^{sf}\bar \eta}{\bar \eta}m_{\perp}(\epsilon)+(1+{\bar \eta}^2)m_{\perp}^{\textrm{sf}}(\epsilon)\right.\\
& \ \left.-
\frac{1}{\bar \eta}(m_{\parallel}(\epsilon)+p^{sf}\frac{\eta^{sf}}{\bar \eta^2}n_e(\epsilon))\right),\\
m_{2 \bfr}(\epsilon)&=-m_{\perp}+2\eta^{\textrm{sf}}m_{1\bfr}(\epsilon),\\
\textrm{with}&\\
m^{\textrm{sf}}_{\perp}(\epsilon)&=
\nu(\epsilon)(\frac{2}{3}\epsilon \partial_{\epsilon} f^{\textrm{sf}}_{\perp}+
f^{\textrm{sf}}_{\perp}),\\
n_{e}(\epsilon)=&\nu(\epsilon)f^0(\epsilon),\\
\ea
\eeq


\begin{thebibliography}{99}

\bibitem{Katine00} J.A. Katine, F.J. Albert, R.A. Buhrman, E.B. Myers
and D.C. Ralph, Phys. Rev. Lett. {\bf 84}, 3149 (2000).

\bibitem{kiselev} S.I. Kiselev, J.C. Sankey, I.N. Krivorotov, N.C. Emley,
R.J. Schoelkopf, R.A. Buhrman and D.C. Ralph,
Nature (London) {\bf  425}, 380 (2003).

\bibitem{Chen04} T.Y. Chen, Y. Ji, C.L. Chien and M.D. Stiles,
Phys. Rev. Lett. {\bf 93} 026601 (2004).

\bibitem{Vernier04} N. Vernier, D.A. Allwood, D. Atkinson, M.D. Cooke
and R.P. Cowburn, Europhys. Lett. {\bf 65} 526 (2004).

\bibitem{Yamaguchi04} A. Yamaguchi, T. Ono, S. Nasu, K. Miyake, K. Mibu
and T. Shinjo, Phys. Rev. Lett. {\bf 92} 077205 (2004).

\bibitem{Klaui05} M. Kl{\" a}ui {\it et al.}, Phys. Rev. Lett. {\bf 95}
026601 (2005).

\bibitem{Hayashi06}  M. Hayashi {\it et al.}, Phys. Rev. Lett. {\bf 96}
197207 (2006).

\bibitem{Thomas06} L. Thomas {\it et al.}, Nature {\bf 443} 197 (2006).

\bibitem{slonczewski} J. C. Slonczewski, J. Magn. Mater.
{\bf 159}, L1 (1996).

\bibitem{berger} L. Berger,
Phys. Rev. {\bf B 54}, 9353 (1996).

\bibitem{zhangli} S. Zhang and Z. Li,
Phys. Rev. Lett. {\bf  93}, 127204 (2004).

\bibitem{bazaliy} Ya.B. Bazaliy, B.A. Jones and S.C. Zhang,
Phys. Rev. {\bf B 57}, R3213 (1998).

\bibitem{barnes} S.E. Barnes and S. Maekawa,
Phys. Rev. Lett. {\bf  95}, 107204 (2005).

\bibitem{thiaville1} A. Thiaville, Y. Nakatani, J. Miltat and Y. Suzuki,
Europhys. Lett. {\bf  69}, 990 (2005).

\bibitem{thiaville2} A. Thiaville, Y. Nakatani, J. Miltat and N. Vernier,
J. Appl. Phys {\bf  95}, 7049 (2004).

\bibitem{lizhang} Z. Li and S. Zhang,
Phys. Rev. Lett. {\bf  92}, 207203 (2004),
{\em ibid} Phys. Rev. {\bf B 70}, 024417 (2004).

\bibitem{he} J. He, Z. Li and S. Zhang,
Phys. Rev. {\bf B 73}, 184408 (2006).

\bibitem{zhang1} S. Zhang, P.M. Levy and A. Fert,
Phys. Rev. Lett. {\bf  88}, 236601 (2002).


\bibitem{review} Y. Tserkovnyak, A. Brataas, G.E.W. Bauer and B.I. Halperin,
Rev. Mod. Phys. {\bf  77}, 1375 (2005).

\bibitem{tatara1} G. Tatara and H. Kohno,
Phys. Rev. Lett. {\bf  92}, 086601 (2004).

\bibitem{waintal} X. Waintal and M. Viret,
Europhys. Lett. {\bf  65}, 427 (2004).

\bibitem{fernandez} J. Fern\'andez-Rossier, M. Braun, A.S. N\'u\~nez and
A.H. MacDonald,
Phys. Rev. {\bf B 69}, 174412 (2004).

\bibitem{xiao} J. Xiao, A. Zangwill and M.D. Stiles,
Phys. Rev. {\bf B 73}, 054428 (2006).

\bibitem{tserkovnyak} Y. Tserkovnyak, H.J. Skadsem, A. Brataas and G.E.W. Bauer,
Phys. Rev. {\bf B 74}, 144405 (2006).

\bibitem{kohno} H. Kohno, G. Tatara and J. Shibata,
cond-mat/0605186 (2006).

\bibitem{sillin} V.P. Sillin,
Zh. Eskp. Teor. Fiz {\bf 33}, 1227 (1957) [Sov. Phys. JETP
{\bf B 6}, 945 (1958)].

\bibitem{white} R.M. White,
{\it Quantum Theory of Magnetism}, (Springer, 1983).
.
\bibitem{mineev} V.P. Mineev,
Phys. Rev. {\bf B 72}, 144418 (2005). {\em ibid} {\bf B 69}, 144429 (2004).

\bibitem{leggett} A.J. Leggett,
J. Phys. C. {\bf 3}, 448 (1970).

\bibitem{spinvalve} A. Shpiro, P.M. Levy, and S. Zhang,
Phys. Rev. {\bf B 67}, 104430 (2003),
 M.D. Stiles, J. Xiao and A. Zangwill
Phys. Rev. {\bf B 69}, 054408 (2004), P.M. Levy, and J. Zhang, Phys.
Rev. {\bf B 70}, 132406 (2004), J. Zhang and P. M. Levy, Phys. Rev.
{\bf B 70}, 184442 (2004), J. Xiao, A. Zangwill and M.D. Stiles,
Phys. Rev. {\bf B 70}, 172405 (2004), J. Zhang and P.M. Levy, Phys.
Rev. {\bf B 71}, 184417 (2005), {\em ibid} {\bf 71}, 184426 (2005), J. Barnas,
A. Fert, M. Gmitra, I. Weymann and  V.K. Dugaev, Phys. Rev. {\bf B
72}, 024426 (2005).

\bibitem{Slonczewski89} J.C. Slonczewski, Phys. Rev. {\bf B 39}
6995 (1989).

\bibitem{Stiles06} M.D. Stiles and J. Miltat, in
{\it Spin Dynamics in Confined Magnetic Structures III}, B.
Hillebrands and A. Thiaville Eds. (Springer, Berlin, 2006).

\bibitem{footnote1} Mathematically,
the extended basis constitutes a linearly independent basis
in the space of adiabatic functions $a(\bfp,\bfr,t)$ defined by 
$a(\bfp,\bfr,t)=
a_{1}\bfn(\bfr,t)+a_{2}\bfn \times \nabla_{\bfr} \bfn+
+a_3 \nabla_{\bfr} \bfn+
a_4 \bfn \times \partial_t \bfn +a_5 \partial_t \bfn$, where the $a_{\alpha}(\bfp)$
are function of $\bfp$ only; i.e $a(\bfp,\bfr,t)=0$ implies  $a_{\alpha}(\bfp)=0$
for all $\alpha$ and $a(\bfp,\bfr,t)=b(\bfp,\bfr,t)$ implies 
$a_{\alpha}(\bfp)=b_{\alpha}(\bfp)$.

\bibitem{generalisation1}
For spin conserving scattering we can generalize our work in two ways:
(i) define distinct intraband
 scattering probabilities $w^{\pm}_{\bfp,\bfp'}$ for energies $\epsilon_{\pm}$.
(ii) define anisotropic scattering probabilities
$w_{\bfp,\bfp'}$.

(i) In establishing the parametric forms $f_{1,2}$ we do
not need the explicit forms of $f^0$ and
$f^0_{\parallel}(\epsilon_{\bfp})$. 
We only need that
$f^0(\epsilon_{\bfp})$ and $f_{\parallel}^0(\epsilon_{\bfp})$
depend only on $\epsilon_{\bfp}$. 
But in fact quite generally we know that 
$f^0(\epsilon_{\bfp})=\frac{1}{2}
(A_+(\epsilon_{\bfp}^+)+A_-(\epsilon_{\bfp}^-))$ and
$f_{\parallel}(\epsilon_{\bfp})= \frac{1}{2} (A_{\parallel
+}(\epsilon_{\bfp}^+)+A_{\parallel -}(\epsilon_{\bfp}^-))$.
Clearly the components $f_{1,2}(\epsilon_{\bfp})$ can also be
split in two components  $A_{1,2 \pm}(\epsilon_{\bfp}^{\pm})$.
Using this property, an immediate generalization of the above
results consists in separating the collision integral into
${\cal I}_{\alpha}={\cal I}_{\alpha}^{++}+{\cal I}_{\alpha}^{--}$
(with obvious notations) and to specify distinct probabilities
$w^{\pm}_{\bfp,\bfp'}= w^{\pm}_{\hat \bfp,\hat
\bfp',\epsilon_{\bfp}^{\pm}}
\delta(\epsilon_{\bfp}^{\pm}-\epsilon_{\bfp'}^{\pm})$. 
In that
way, for isotropic scattering, all the previous expressions are
valid when interpreted for each index $\pm$, and this allows
considering the possibility of two distinct scattering times
$\tau_{\pm}$.

(ii)  Anisotropic scattering consists in writing
$w_{\bfp,\bfp'}=\sum_{\ell} w^{\ell}_{\epsilon_{\bfp}}
P_{\ell}(\hat \bfp\cdot \hat \bfp')$ where
$P_{\ell}(\hat \bfp\cdot \hat \bfp')$ are
Legendre polynomials of order $\ell$.
In the equilibrium situation, the only modification of our results consists
in replacing the isotropic scattering time
$1/\tau=w^0 (\epsilon)\nu(\epsilon)$ by the transport scattering time
$1/\tau_{\textrm{tr}}=(w^0 (\epsilon)-\frac{1}{3}w^1 (\epsilon))\nu(\epsilon)$.
In the out-of-equilibrium situation the analysis becomes more subtle and is beyond the
scope of the present work.

\bibitem{inpreparation} F. Pi{\'{e}}chon and A. Thiaville, in preparation.

\bibitem{antropov} J. Zhang, P.M. Levy, S. Zhang and V. Antropov
Phys. Rev. Lett. {\bf 93}, 256602 (2004), {\em ibid} {\bf 96}, 019708 (2006) and
J.C. Slonczewski Phys. Rev. Lett. {\bf 96}, 019707 (2006).

\bibitem{remarks1} It appears in fact that, to describe a
time dependent effect to linear order, it is necessary to consider
an extended basis of the nine following vectors
$\bfn_L,\bfn_R,\bfn_L \times \bfn_R$ and the six new vectors
$\partial_t \bfn_R,\bfn_L\times \partial_t \bfn_R, \bfn_R\times
\partial_t \bfn_R, \bfn_L\times(\bfn_L\times \partial_t \bfn_R),
\bfn_L\times(\bfn_R\times \partial_t \bfn_R) $ and
$\bfn_L\times(\bfn_L\times(\bfn_R\times \partial_t \bfn_R))$.
In this basis the mathematical properties described in \cite{footnote1} are valid.

\end{thebibliography}
\end{document}